
\nopagenumbers
\magnification=\magstep1
\parskip 0pt
\parindent 15pt
\baselineskip 16pt
\hsize 5.53 truein
\vsize 8.5 truein

\font\titolo = cmbx10 scaled \magstep2
\font\autori = cmsl10 scaled \magstep2
\font\abstract = cmr9

\def\CC {{\hbox{\tenrm C\kern-.45em{\vrule height.67em width0.08em
        depth-.04em \hskip.45em}}}}
\def\DD {{\hbox{\tenrm {I\kern-.18em{D}}\kern-.36em {\vrule
        height.62em width0.08em depth-.04em\hskip.36em}}}}

\def\RR {{\hbox{\tenrm I\kern-.17em{R}}}}
\def\BB {{\hbox{\tenrm I\kern-.17em{B}}}}
\def\ZZ {{\hbox{\tenrm Z\kern-.31em{Z}}}}
\def\ZB {{\hbox{\tenrm Z\kern-.31em{Z}$_2$}}}
\def\KK {{\hbox{\tenrm I\kern-.17em{K}}}}
\def\efi{{\hbox{$e^{i \psi}$}}}
\def\efis{{\hbox{$e^{-i \psi}$}}}
\def\olf {{\hbox{\tensy O\kern-.68em{O}$_f$}}}
\def\olfa {{\hbox{\tensy O\kern-.68em{O}$_{f_1}$}}}
\def\olh {{\hbox{\tensy O\kern-.68em{O}$_h$}}}
\def\olx {{\hbox{\tensy O\kern-.68em{O}$_x$}}}
\def\gru{{\hbox{$G$}}}
\def\grh{{\hbox{$H$}}}
\def\grw{{\hbox{$W$}}}

\def\alg{{\hbox{\tensy G}}}
\def\alga{{\hbox{\tensy G$_0$}}}
\def\algb{{\hbox{\tensy G$_1$}}}
\def\algab{{\hbox{$\alga \oplus \algb$}}}
\def\algh{{\hbox{\tensy H}}}
\def\algl{{\hbox{\tensy G$_\Lambda$}}}
\def\algn{{\hbox{\tensy N}}}
\def\algs{{\hbox{\tensy G$^\star$}}}
\def\algls{{\hbox{\tensy G$_\Lambda^\star$}}}
\def\algw{{\hbox{\tensy W}}}
\def\algwl{{\hbox{\tensy W$_\Lambda$}}}
\def\algws{{\hbox{\tensy W$^\star$}}}
\def\bbf{{\hbox{$\omega_f$}}}
\def\ccf{{\hbox{$\varpi_f$}}}
\def\grass{{\hbox{$\Lambda$}}}
\def\graa{{\hbox{$\Lambda_0$}}}
\def\grab{{\hbox{$\Lambda_1$}}}
\def\graab{{\hbox{$\graa \oplus \grab $}}}
\def\grado{{\hbox{\sevenrm deg}}}
\def\gsh {{\hbox{\tensy M\kern-1.14em{M}}}}
\def\id {{\hbox{\tenrm I\kern-.19em{I}}}}
\def\jj {{\hbox{\tenrm J\kern-.35em{J}}}}
\def\ko{{\hbox{$ K_0(g)$}}}

\def\a{{\hbox{$a$}}}
\def\b{{\hbox{$b$}}}
\def\as{{\hbox{$a^\dagger$}}}
\def\bs{{\hbox{$b^\dagger$}}}

\def\cona{{\hbox{\tenrm Con}}}
\def\conb{{\hbox{\tenrm ecture:}}}
\def\conj{{\hbox{$\underline{\cona}$j$\underline{\conb}$}}}

\def\qq {{\hbox{\tentt [\kern-0.07em{]}}}}

\def\cfr#1{{\hbox{$^{#1}$}}}
\def\nr{{\hbox{3}}}

\def\cc {{\hbox{\rm C\kern-.45em{\vrule height.67em width.08em depth-.04em
\hskip.45em}}}}
\def\ii {\hbox{\rm I\kern-.19em{I}}}
\def\zz {\hbox{\rm Z\kern-.35em{Z}}}
\def\rr {\hbox{\rm I\kern-.17em{R}}}
\def\mm{{\hbox{\tensy M\kern-1.14em{M}}}}
\def\hh {{\hbox{\tensy H\kern-.78em{H}}}}
\def\jj {{\hbox{\tenrm J\kern-.35em{J}}}}

\def\uu{{\hbox{$u(1|1)$}}}
\def\uul{{\hbox{$u(1|1)_\Lambda$}}}
\def\uuls{{\hbox{$u(1|1)_\Lambda^\star$}}}

\def\su{{\hbox{$su(2|1)$}}}
\def\sul{{\hbox{$su(2|1)_\Lambda$}}}
\def\suls{{\hbox{$su(2|1)_\Lambda^\star$}}}
\def\SU{{\hbox{$SU(2|1)$}}}

\def\uo{{\hbox{$uosp(1|2)$}}}
\def\uol{{\hbox{$uosp(1|2)_\Lambda$}}}
\def\uols{{\hbox{$uosp(1|2)_\Lambda^\star$}}}

\hrule height 0pt
\vskip .85truein
\hrule height 0pt
\vskip 1.5 truein
\centerline{\titolo GENERALIZED COHERENT STATES}
\vskip .05truein
\centerline{\titolo FOR DYNAMICAL SUPERALGEBRAS}
\vskip .15truein
\centerline{{\autori Alessandro Pelizzola}\footnote{$^\flat$}{\abstract
Dipartimento di Fisica, Politecnico di Torino, 10129 Torino, Italy}
{\autori and Corrado Topi}$^{\,\flat}$}
\vskip .35truein
\moveright 1.5 truein \vbox{\hrule width 2.9 truein height 1pt}
\vskip .35truein
\centerline{\bf ABSTRACT}
\vskip .05truein {\abstract Coherent states for a general Lie superalgebra
are defined following the method originally proposed by Perelomov.
Algebraic and geometrical
properties of the systems of states thus obtained are examined, with
particular attention to the possibility of defining a K\"ahler structure
over the states supermanifold and to the connection between this
supermanifold and the coadjoint orbits of the dynamical supergroup.
The theory is then applied to some compact forms of contragradient Lie
superalgebras.}
\vfill
 \par
\eject

\pageno=1
\footline{\hss\tenrm\folio$\,$\hss}
\def\makefootline{\baselineskip=50pt\line{\the\footline}}

\centerline{\bf 1. INTRODUCTION}
\smallskip
Coherent states (CS) have been first introduced in quantum optics by
Glauber$^{1,2}$ as the states of minimum quantum uncertainty for
the harmonic oscillator. Glauber's CS preserve their shape
during time evolution. After the work of Glauber several attempts
have been made to generalize the concept of CS to systems with a larger
dynamical group$^{3,4}$. The most promising among them is certainly
the method proposed by Perelomov$^5$.
Perelomov defined generalized coherent states
(GCS) for an arbitrary Lie group by mimicking the group-theoretical
properties of Glauber's CS. GCS for a general dynamical system fulfill
three requirements: a) they are states of minimum quantum uncertainty, b)
their manifold is just the classical phase space of the system; c) the
quantum evolution of the system can be described by a path on this
manifold, that is the solution of Euler-Lagrange's equations. \par
In the past few years several papers appeared in which the concept of coherence
is carried over to systems that exhibit a dynamical symmetry with an infinite
number of degrees of freedom$^6$ and to supersymmetric systems.
Supercoherent states (SCS) have been defined for series of noncompact
orthosymplectic supergroups$^{7,8,9}$. A definition that already
contains in itself the general idea is given in Ref. 10 in the context of a
condensed matter problem, while the problem of studying SCS in their full
generality is first approached by the authors$^{\,11,12}$, with a
specific attention to the phase space properties of the SCS system, and by
Nieto and co-workers$^{13}$, whose examples fulfill the requirement
of minimum quantum uncertainty. \par
The purpose of the present paper is to define SCS associated to a general
Lie superalgebra, namely to a graded algebra of even (bosonic) and odd
(fermionic) operators that satisfy certain commutation and anticommutation
rules.
Several quantum mechanical
systems are described by a second-quantized hamiltonian that lives in a
dynamical algebra which is in fact a Lie superalgebra. The (super-) algebra
${\cal A}$ is said to be a dynamical (super-) algebra for the hamiltonian $H$
if the operators that appear in $H$ close by commutation
(and/or anticommutation) on a set of generators of
${\cal A}$.  \par
Dynamical superalgebras enter quantum mechanics in different ways. A very
interesting instance is
the fermionic linearization of many fermion systems second quantized
hamiltonians$^{14}$. A product of two fermionic operators $A$ and
$B$ describing different modes of the system can be written, under the
condition
$(A - \langle A \rangle) (B - \langle B \rangle) \simeq 0$, in the form
$$AB \simeq \langle A \rangle B + A \langle B \rangle -
\langle A \rangle \langle B \rangle \eqno(1.1)$$
that can be expected to be fulfilled in quasiclassical states, as SCS must be.
Since $A$
and $B$ are supposed to belong to different modes, they must satisfy
$\{A,B\} = 0$, and, for the consistency of eq. (1.1), $\langle A
\rangle$ and $\langle B \rangle$ (which could for example be the expectation
values self consistently evaluated in the ground state) must be taken as odd
elements of a Grassmann algebra \grass (a-numbers), anticommuting among
themselves and with the fermionic operators. The requirement of
anticommutation between fermionic operators and a-numbers can be relaxed
(and in fact we relax it) by writing eq. (1.1) in the form
$$AB \simeq \langle A \rangle B - \langle B \rangle A -
\langle A \rangle \langle B \rangle, \qquad \langle A \rangle , \langle B
\rangle \in \grass .$$\par
The paper is organized as follows.
In Ch. 2 we review the theory of GCS, give the basic definitions for SCS
mimicking those of GCS and
study their algebraic properties as well as the geometrical structure of
their supermanifold. In Ch. 3 we examine some properties of the coadjoint
orbits of Lie supergroups. The main objectives
are to determine whether the supermanifold is K\"ahler (and then can be
interpreted as a classical phase space) or not and to
establish the connection between SCS supermanifolds and coadjoint orbits of
the dynamical supergroup. We believe that the coadjoint orbits approach
should be a powerful tool to study all the SCS systems associated to a
certain dynamical superalgebra, especially when the latter is nilpotent. \par
In Ch. 4 some examples are studied, in which the dynamical superalgebra is
a compact form of a contragradient Lie superalgebra, and, finally,  in Ch. 5 we
summarize our results and propose a conjecture concerning the realizability
of SCS systems whose manifolds are endowed with K\"ahlerian structure. \par
\eject
\centerline{\bf 2. GENERALIZED COHERENT STATES}
\centerline{\bf FOR DYNAMICAL SUPERALGEBRAS}
\medskip
\bf 2.1 Definitions. General properties \par
\smallskip
\rm
First of all, we shall briefly review the fundamental definitions and
properties of the theory of generalized coherent states (GCS) for ordinary
(dynamical) Lie algebras \cfr{\,5,15}. \par
Let ${\cal G}$ be the dynamical algebra of our model, $G$ an element
of the class of the corresponding Lie
groups and $T$ a unitary irreducible representation
(UIR) of $G$ in a Hilbert space $V$. \par
We assume that in the representation space $V$ there exists a fixed cyclic
vector, that we shall denote by $|\psi_0\rangle$,
and call $H$ the set of elements $h \in G$ such that
$$T(h)|\psi_0\rangle  = e^{i\alpha(h)}|\psi_0\rangle , \qquad
\alpha : H \rightarrow \rr. $$
It is easy to verify that $H$ is a subgroup of $G$, which will be called
isotropy subgroup of $|\psi_0\rangle $
and that $e^{i\alpha}$ must be a unitary character of $H$. \par
Let $\mm=G/H$ be the left coset of $G$ with respect to $H$. We see that
there is no arbitrariness in the choice of a particular group $G$
among all those associated with the algebra ${\cal G}$, because the
manifold $\mm$ depends only on ${\cal G}$. \par
Now we can define the coherent states of ${\cal G}$ by means of a mapping
from $\gsh$ to $V$, which associates to each $x \in \mm$ defined by the
decomposition $g = x \cdot h$, with $g \in G$ and $h \in H$, the state (up
to a phase factor)
$$|x\rangle  = T(x)|\psi_0\rangle . $$
Thus the coherent states are represented by the points of a manifold $\gsh$
that will be called the coherent states
manifold, on which $G$ acts transitively by means of the left translation
$\circ : G \times \mm \rightarrow \mm$ defined by
$$g \circ x = \pi \left( g \cdot \pi^{-1}(x) \right), \qquad
\forall g \in G, \, x \in M, \eqno(2.1)$$
$\pi$ being the natural projection of $G$ on $\mm$, $G$ being
seen as a bundle over
$\mm$. The transitivity of \gsh with respect to $G$ reflects the first
important algebraic property of the system of states just defined: the
action of the dynamical group $G$ maps coherent states into other
coherent states. \par
Another algebraic property is the resolution of identity, of
fundamental importance in the coherent states representation of the Feynman
path integral$^{16}$. Let $d\mu(x)$, $\forall x \in \mm$ be a $G$-invariant
measure on the coherent states manifold (which can be deduced either from an
invariant metric or, in the simplest cases, from the invariant Haar measure on
the group itself). When the integral converges, the operator
$$B = \int d\mu(x)|x\rangle \langle x| \eqno(2.2)$$
commutes with all the operators of the representation $T$ (namely
it is invariant
with respect to $T$):
$$T(g)BT(g)^{-1} = B, \qquad \forall g \in G,\eqno(2.3)$$
and, by Schur's lemma, is a multiple of the identity, $B = k\ii $, hence
$${1 \over k}\int d\mu(x) |x\rangle \langle x| = \ii. \eqno(2.4)$$
Notice that $k$ can be obtained from the relation
$$k = \langle y|B|y\rangle  = \int d\mu(x) |\langle y|x\rangle |^2,
\eqno(2.5)$$
where $|y\rangle  \in V$ is normalized. \par
A consequence of this property is the completeness of the system of
coherent states: an arbitrary state $|\psi\rangle  \in V$ can, in fact, be
expanded in terms of coherent states
$$|\psi\rangle  = {1 \over k} \int d\mu(x) \psi(x) |x\rangle ,
\qquad \psi(x) = \langle x|\psi\rangle.\eqno(2.6)$$
The function $\psi(x)$ is a solution of the integral equation
$$\psi(x) = {1 \over k} \int d\mu(y) \langle x|y\rangle  \psi(y), \eqno(2.7)$$
where the kernel $K(x,y) = {\displaystyle {1 \over k}}\langle x|y\rangle $
is self-reproducing:
$$K(x,z) = \int d\mu(y) K(x,y) K(y,z). \eqno(2.8)$$
Indeed, the system of coherent states is overcomplete. One of the features
whereby this property can be derived is its cardinality, that is continuum.
\par
Now, we consider instead of an ordinary Lie
algebra ${\cal G}$, a Lie superalgebra, that we shall still denote with
${\cal G}$. A Lie superalgebra is a $\ZB-$graded space ${\cal G}$,
i.e. a vector space which is the direct sum of two vector subspaces
${\cal G}_0$ and
${\cal G}_1$, endowed with an
algebraic structure denoted by $ [,] $, obeying the axioms
$$\eqalign{[a,b]= & -(-1)^{\grado(a) \grado(b)}[b,a] \cr
[a,[b,c]] = & [[a,b],c] + (-1)^{\grado(a)\grado(b)}[b,[a,c]] \cr}$$
where deg is a linear mapping ${\rm deg}:\alg \rightarrow \ZB $ called
parity obeying the property
$$ {\rm deg}:a \longmapsto \alpha, \qquad \forall a \in \alg_\alpha, \quad
\alpha \in \ZB. $$
We shall use in the following the definition
$$[a,b] = ab - (-1)^{\grado(a) \grado(b)} ba.$$
 \par
The group associated to a Lie superalgebra by exponential mapping is a Lie
supergroup$^{17,18}$, somewhat similar to a Lie group with parameters that
take values in the even or odd (depending on the parity of the
corresponding generator in the Lie superalgebra) subspace of a Grassmann
algebra $\grass .$ \par
Once more, we consider a UIR of the supergroup $G$ in a
$\zz_2$-graded Hilbert space $V$, fix a cyclic vector $|\psi_0\rangle$ and
proceed as before. The left coset space $\mm = G/H$ will be a
supermanifold$^{17}$
(that is, a manifold with even and odd coordinates), and coherent states (now
SCS) may be defined in the usual way. However, two points are worthy of
enlightenment about the properties previously demonstrated for GCS.
The integration in (2.2)-(2.8) must be intended as an ordinary integration
for what concerns
even variables, and as a Berezin integration$^{19}$ for what concerns odd
variables. We recall that if $\eta$ is a variable taking values in the odd
subspace of a Grassmann algebra, Berezin integral is defined by
$$\int d\eta = 0 \qquad \int \eta \, d\eta = 1. $$
This is a formal definition that can be interpreted as
a definite integral over the whole domain of $\eta$ but does not involve any
concept of measure theory, thus we do not have notions of indefinite
integral, or of integrals over subdomains. We will come back later to
this problem, when talking about Berry's phase. \par
In the same way, the invariant measure $d\mu(x)$ must be intended as
invariant with respect to the integration just defined, and it will not
therefore
be a measure function, at least for its odd part. Given the
$G$-invariant metrics $g$ that we will construct in all our examples, the
invariant measure can be obtained from the relation
$$d\mu(x) = |{\rm sdet}[g(x)]|^{1/2}, \eqno(2.9)$$
where sdet denotes the superdeterminant.
\footnote{$^{(a)}$}{We recall that given a $(p + q) \times (p + q)$
supermatrix $M = \pmatrix{A & B \cr C & D \cr}$, with $A$ and $D$
respectively $p \times p$ and $q \times q$ matrices with even entries, and
$B$ and $C$ $p \times q$ and $q \times p$ matrices with odd entries, one
defines the superdeterminant
${\rm sdet}M = {\rm det}(A - BD^{-1}C) {\rm det}D^{-1}$ and the supertrace
${\rm str}M = {\rm tr}A - {\rm tr}D$.}
As for the resolution of identity,
eq. (2.3) is still valid, but it is not sufficient to ensure the validity
of (2.4),
because Schur's lemma generalizes to the graded case with two alternatives.
$B$ can be, in fact, either a multiple of the identity or an operator that
permutes
the homogeneous subspaces of $V$, when they have the same dimension. In present
case, however, $B$ is even, and the second possibility has to be excluded. \par
\medskip
\bf 2.2 The case of contragradient Lie superalgebras \par
\smallskip
\rm
In this section contragradient Lie superalgebras are to be intended in the
sense of Ref. 20. They are Lie superalgebras for which a Cartan-Weyl
basis can be defined. They are particularly interesting in present context,
and we restrict in this section our attention to them, because of the
features of their irreducible representations theory, that are very similar to
those of semisimple Lie algebras. \par
Actually, our examples will deal with compact forms of: \parindent 0pt \par
a) contragradient Lie superalgebras modulo their centers, \par
b) Cartan extensions of contragradient Lie superalgebras. \parindent 15pt \par
The most important result established by Kac in this sense is that
the irreducible representations of these superalgebras admit a
highest weight vector in the representation space. In order to clarify why
this is important for our purposes we must go back to the theory of
GCS for ordinary Lie algebras. GCS are actually "coherent", that is, they are
states of minimum quantum uncertainty: in fact, it was shown in Ref. 21 that
GCS for compact semisimple Lie algebras minimize the variance of the quadratic
Casimir operator, which is the simplest invariant estimate of quantum
uncertainty, when they are constructed choosing as base  the highest
weight vector $|\psi_0\rangle$ of the representation $T$. In the book by
Perelomov$^{22}$ this
result is extended to other algebras (Weyl-Heisenberg, $su(1,1)$) by
requiring that $|\psi_0\rangle $ is choosen in such a way that its isotropy
subalgebra is maximal. The isotropy subalgebra ${\cal B}$ of a vector
$|\psi_0\rangle$ is the set of elements $t $ of the complex extension
$ {\cal G}^c = {\cal G} \oplus
i{\cal G}$ which satisfy
$$t|\psi_0\rangle  = \lambda_t|\psi_0\rangle , \qquad \lambda_t \in
\cc,$$
and is said to be maximal when ${\cal B} \oplus \bar{\cal B} = {\cal G}^c$,
where $ \bar{\cal B}$ is the algebra of the hermitian conjugates of the
elements belonging to ${\cal B}$.
If ${\cal G}$ is semisimple and $|\psi_0\rangle $ is the highest weight vector
in representation space, then ${\cal B}$ is the subalgebra of ${\cal G}^c$
generated by the Cartan and the raising operators
$${\cal B} = {\cal H} \oplus \sum_{\alpha \in \Delta_+} {\cal G}_\alpha,
\eqno(2.10)$$
for which, of course, ${\cal B} \oplus \bar{\cal B} = {\cal G}^c$. \par
The relation (2.10) still holds for contragradient Lie superalgebras (now
raising operators can be either bosonic or fermionic; slight modifications
are to be made to include the case of degenerate representations, see Sec.
2.4) thus we propose to define SCS
for contragradient Lie superalgebras using the highest weight vector as base
vector $|\psi_0\rangle $. \par
Besides, we must observe that this no longer implies "coherence", in the
sense of Delbourgo \cfr{21},
because his proof relies on definite positivity of the
quadratic Casimir operator, which no longer holds in general in the case of Lie
superalgebras. \par
For a real semisimple Lie algebra the algebra of the isotropy subgroup $H$
(that is not the isotropy subalgebra ${\cal B}$) is
just the Cartan subalgebra ${\cal H}$ if the representation is
non-degenerate, but can be larger if the representation is degenerate.
The same happens for dynamical superalgebras.
There are non-degenerate representations
in which the highest weight vector is
annihilated only by the raising operators: in this case the algebra of $H$
is again the Cartan subalgebra, and then $H$ is an ordinary Lie group, the
Cartan subgroup. This allows us to decompose the SCS supermanifold into the
(in general semidirect) product $G/H = G/G_0 \otimes_s G_0/H$, where
$G_0$ is
the ordinary Lie group associated to the even subalgebra ${\cal G}_0$.
Notice that the factor $G/G_0$ is purely fermionic (it has only odd
coordinates), while the factor $G_0/H$ is purely bosonic (it is just the
GCS manifold of the Lie group $G_0$). A consequence of this property is
that the invariant measure (2.9) can be written as a product of two
functions, one depending only on the odd coordinates, and the other on the
even ones.
There are also degenerate representations (see the
example of $su(2|1)$) in which the highest weight vector is annihilated also
by some (bosonic and/or fermionic) lowering operators:
then these operators, and their conjugates,
belong to the algebra of $H$ and thus appear as generators of $H$, which may
then become a true Lie supergroup. \par
\medskip
\bf 2.3 Stability \par
\smallskip
\rm
The concept of stability of coherent states is intimately related to that
of dynamical (super-) algebra associated to an hamiltonian.
By virtue of the definition of dynamical (super-) algebra given in the
Introduction, the time-evolution operator, that can be obtained for
example by Magnus' formula$^{23}$, is an element of the exponential group $G =
\exp {\cal G}$ associated with the dynamical algebra (the dynamical group),
that is, $U(t,t_0) = \exp \gamma$, for some $\gamma \in {\cal G}$ and, by
definition of coherent states, it will be (from now on, we shall write group
and algebra elements instead of their representatives)
$$U(t,t_0) |z_0\rangle = \exp \gamma \cdot \exp Z_0 |\psi_0\rangle =
\exp Z |\psi_0\rangle = e^{i\phi} |z\rangle, $$
with $|z_0\rangle$ and $|z\rangle$ coherent states and $Z_0,Z \in
{\cal G}$. \par
As a consequence, a system that has been prepared in a coherent state
$|z_0\rangle$ at the time $t_0$ will be found in any other time in a
coherent state $|z\rangle$ still.\par
Actually, if ${\cal G}$ is not semisimple, the "coherence-preserving"
hamiltonians may live in a larger algebra ${\cal S}$ in the universal
enveloping algebra of ${\cal G}$, that contains ${\cal G}$ as an
ideal$^{24}$. If we are dealing with ordinary Lie
algebras (but we assume this result to be valid also for Lie superalgebras)
${\cal S}$ is the algebra of the automorphism group of ${\cal G}$, and when
${\cal G}$ is solvable ${\cal S}$ is the algebra with Levi decomposition
${\cal S} = {\cal A} \oplus {\cal G}$, with ${\cal G}$ maximal solvable
ideal of ${\cal S}$. \par
\medskip
\bf 2.4 Coherent states supermanifold \par
\smallskip
\rm
Ordinary GCS are parametrized by the points of a complex manifold $\mm$ (e.g.
the complex plane for Glauber coherent states, the 2-dimensional sphere for
SU(2)-coherent states). It is very interesting to ask whether this manifold
is a $G$-homogeneous K\"ahler manifold$^{25}$ or not (the answer for both
Glauber and SU(2) coherent states is positive). This problem
has been completely solved
for compact semisimple Lie algebras in Ref. 26. \par
If it is, there exists on $\mm$ a $G$-invariant metric $g$ and a
metric-preserving complex structure \jj, namely there exists
$\jj: T_m\mm \rightarrow T_m\mm$, where $T_m$ is the space tangent to $\mm$
in the point $m$, such that
$$\jj^2 = - \id, \qquad g(\jj x,\jj y) = g(x,y) \quad \forall m \in \mm, \,
\forall x,y \in T_m\mm. \eqno(2.11)$$
The metric and the complex structure can be used to define a two-form
$\omega$ by
$$\omega(x,y) = g(\jj x,y). $$
The manifold $\mm$ is said to be K\"ahler if such two-form is closed
($d\omega = 0$). Then $\mm$ is also symplectic and can be viewed as a classical
phase space for our dynamical system. This is easy exemplified in the
case of Glauber coherent states by the identification
$$a = {q + ip \over \sqrt{2\hbar}}, \qquad
a^\dagger = {q - ip \over \sqrt{2\hbar}}. $$
Let us see how coherent states allow us to map a quantum-mechanical
problem into a classical dynamical one on more general grounds. First of all
we recall the notion of K\"ahler potential. Let us introduce in $\mm$
an atlas of local coordinate system $\left\{ z^i, \bar z^i \right\}$
such that $\displaystyle
{\partial \over \partial z^i}, {\partial \over \partial \bar z^i}$ are
eigenvectors of $\jj$:
$$\jj{\partial \over \partial z^i} = i {\partial \over \partial z^i}, \qquad
\jj{\partial \over \partial \bar z^i} = -i {\partial \over \partial \bar z^i}
.\eqno(2.12)$$
Then we define the components of the metric
$$g_{j \bar i} = g \left( {\partial \over \partial z^j}, {\partial \over
\partial \bar z^i} \right), $$
while
$$g \left( {\partial \over \partial z^i}, {\partial \over \partial z^j}
\right) = g \left( {\partial \over \partial \bar z^i}, {\partial \over
\partial \bar z^j} \right) = 0 $$
because of (2.11) and (2.12). When the two-form $\omega$ is closed (i.e.
when the manifold is K\"ahler) there exists a function $K(z^i, \bar z^i)$
from which the metric can be obtained by derivation
$$g_{i \bar j} = {\partial^2 K \over \partial z^i \partial \bar z^j}.
$$
The function $K$ is just the K\"ahler potential. \par
The K\"ahler potential plays a crucial role in the dynamics on the coherent
states manifold. We have already showed that coherent states are defined in
such a way that the time-evolution operator maps a coherent state into
another one: it follows that (up to a phase factor) all the dynamical
problem is reduced to the determination of a path in the coherent states
manifold, and the system is described by a state
$$|\psi(t)\rangle  = e^{i\alpha(t)}|z(t)\rangle  $$
at any time $t$. \par
The phase factor $e^{i\alpha(t)}$ is indeed very important; the phase
$\alpha(t)$ is but the effective action that appears in a functional
integral approach$^{27}$ and its form is as follows$^{28}$
$$\alpha(t) = - \int_0^t h(z(\tau), \bar z(\tau)) d\tau +
{\rm Im} \int_0^{z(t)} {\partial K \over \partial z} dz, $$
where $h(z, \bar z) = \langle z|\hh|z\rangle $, $\hh$ denoting the
hamiltonian operator, and $K \equiv K(z, \bar z)$ is the
K\"ahler potential of $\mm$. The second term is a geometrical one and can be
viewed as a Berry's phase$^{29}$. It has been conjectured \cfr{30}
that in the context of
the models for High-$T_c$ superconductivity, this term could represent the
element bridging Hubbard-like hamiltonians$^{31}$ with
Chern-Simons field theories$^{32}$, and therefore possibly with anyons. \par
We have now an additional problem with respect to the customary Lie algebra
case. In fact the second
integral is not at all well-defined. This is why it is conjectured that Berezin
integration is a too formal one for our purposes and the field of a-numbers
might be different from pure Grassmann\cfr{33}.\par
Let us now return to the problem of dynamical
superalgebras and study the SCS supermanifold that arises in
this case. \par
If we restrict our attention to contragradient Lie superalgebras we can
still say that the SCS supermanifold is complex, because
SCS are of the form
$$|x\rangle  = \exp \left\{ \sum_{\alpha \in \Delta_0^+} (z^\alpha E_\alpha -
{z^\alpha}^* E_{-\alpha}) + \sum_{\alpha \in \Delta_1^+ {\rm mod} h_1}
(\zeta^\alpha E_\alpha - {\zeta^\alpha}^* E_{-\alpha}) \right\} |\psi_0\rangle
,\eqno(2.13)$$
where $\Delta_\alpha^+$ is the set of positive roots of grade $\alpha$,
$\alpha \in \zz_2$;
$h_1$ is the set of positive fermionic roots that enter the isotropy subalgebra
of $|\psi_0\rangle $ when the representation is degenerate (see Sec. 2.2);
$E_{-\alpha} = E_\alpha^\dagger$, with $E_\alpha$ ladder operator
corresponding to the root $\alpha$;
$z^\alpha \in \Lambda_0$ and $\zeta^\alpha \in \Lambda_1$ ($\Lambda_0$ and
$\Lambda_1$ denote respectively
even and odd subspaces of a Grassmann
algebra $\Lambda$ over the complex field), and
$^*$ is the ordinary conjugation in $\grass$ that satisfies:
$${\rm deg}(\lambda^*) = {\rm deg}(\lambda), \qquad (\lambda^*)^* =
\lambda, \qquad (\lambda\mu)^* = \mu^* \lambda^*.$$
In the following, we shall denote by $\rr_c$ ($\rr_a$) the set of elements
$\lambda \in \graa$ ($\grab$) for which $\lambda^* = \lambda$.
There will be a sign and notation change (see Sec. 4.3)
when the representation is a grade star$^{34}$ one. \par
The relevant question is again whether $\mm$ is a K\"ahler supermanifold
or not. \par
In Ch. 4 we shall examine four examples, and
derive grounds for a conjecture, presented in the Conclusions
that applies to a general theory. In the examples we
shall explicitly realize the following general scheme \cfr{35}.
Let ${\cal G}$ denote the dynamical superalgebra, ${\cal H}$ the algebra of
the isotropy subgroup of $|\psi_0\rangle $ and ${\cal M}$ the complementary
subspace of ${\cal H}$ in ${\cal G}$ (this is not an algebra)
$${\cal G} = {\cal H} \oplus {\cal M}. $$
Introducing in ${\cal G}$ a basis of homogeneous generators $\{X_\alpha\}$
such that ${\cal M}$ is generated by a subset $\{X_\mu\}$, the coset
representatives will be written as
$$x = \exp \left\{ \sum_{X_\mu \in {\cal M}} x^\mu X_\mu \right\},
$$
where the coordinates $x^\mu$ of $x$ will be c-numbers or a-numbers (that
is, even or odd elements of the Grassmann algebra) depending on
the parity of $X_\mu$. \par
In order to construct a $G$-invariant metric on $\mm$ we determine the action
of the group $G$ on the manifold $\mm$ by means of the equation
$$\delta g \cdot x = x^\prime \cdot \delta h, \eqno(2.14)$$
where
$$\delta g = \exp \left\{ \sum_{X_\alpha \in {\cal G}} g^\alpha X_\alpha dt
\right\} $$
is a group element close to the identity, $x$ (given by the (2.13)) and
$$x^\prime = \exp \left\{ \sum_{X_\mu \in {\cal M}} (x^\mu + dx^\mu) X_\mu
\right\} $$
are coset representatives, and
$$\delta h = \exp \left\{ \sum_{X_\alpha \in {\cal H}} dh^\alpha X_\alpha
\right\} $$
is an element of the isotropy subgroup close to the identity. \par
Notice that equation (2.14) is but the infinitesimal version of
(2.1). It is the matrix form of a linear system in the
$\left\{ dx^\mu, X_\mu \in {\cal M} \right\}$'s whose solutions
have the form
$$dx^\mu = \sum_{X_\alpha \in {\cal G}} g^\alpha \sigma^\mu_\alpha(x) dt,
\qquad X_\mu \in {\cal M}, $$
or, in vector notation,
$$dx = g^\alpha \sigma_\alpha dt. $$
A $G$-invariant metric must have the $\sigma_\alpha$'s as Killing vector
fields (generators of isometries):
$$\ell_{\sigma_\alpha}g = 0 \qquad \forall X_\alpha \in {\cal G}
$$
($g$ denotes the metric and $\ell$ Lie derivation) and these equations can
be solved if the superalgebra ${\cal G}$ admits a nondegenerate, invariant
and supersymmetric bilinear form $\gamma$. The properties of invariance and
supersymmetry are fulfilled by the Cartan-Killing form $\gamma_{CK}(X,Y) =
{\rm str}({\rm ad}X{\rm ad}Y)$, with $({\rm ad}X)(Y) = [X,Y]$ where the
symbol $[,]$ is the graded commutator, that can be identified with
$\gamma$ when it is
nondegenerate. In other cases one must construct alternative forms. \par
The metric $g$ must then be taken equal to
$\gamma$ at some fixed point (e.g. $x^\mu = 0, \forall X_\mu \in {\cal M}$)
and "moved" to any other point by means of the $\sigma_\alpha$'s. The
inverse of the metric so obtained is given by
$$g^{\mu \nu}(x) = (-)^{\alpha(\mu + 1)} \sigma^\mu_\alpha(x)
\sigma^\nu_\beta(x) \gamma^{\alpha\beta}, \eqno(2.15)$$
where $(-)^{\alpha(\mu + 1)}$ is a shorthand for
$(-1)^{{\rm deg}(X_\alpha)[{\rm deg}(X_\mu) + 1]}$. \par
So far we have obtained a $G$-invariant metric on the complex supermanifold
$\mm$, but we still cannot say whether the supermanifold is K\"ahler or not.
The
answer to this question requires the determination of the complex
structures $\jj$ and their eigenvector basis in $T_m\mm$. This (difficult)
passage gives rise to very cumbersome forms, and so
(with the exception of the example of
$u(1|1)$, in which the metric given by (2.15) is already K\"ahler) we shall
not give it in explicit detail in the
examples with a larger dynamical superalgebra ($su(2|1)$ and $uosp(1|2)$). \par
\bigskip
\centerline{\bf 3. COADJOINT ORBITS APPROACH}
\centerline{\bf TO COHERENT STATES PROBLEM.}
\medskip
{\bf 3.1 A review of the ordinary formulation.}
\smallskip
In this section we shall describe the main tool of the coadjoint orbits
theory, following the original formulation of Kirillov\cfr{36,37} and
demonstrate the
connection with coherent states problem for nilpotent Lie groups, according to
the approach proposed by Moscovici\cfr{38}.

Let $\alg$ be the ordinary finite dimensional Lie algebra.
Since $\alg$ is a vector space, we can construct on $\alg$ the space
$\algs$ of exterior 1-forms, which is isomorphic to \alg$.$ \par
Let $G$ be the corresponding Lie group. We call the coadjoint representation of
$\gru,$ the representation
of the group that we can construct on $\algs,$ by means of the relation
$$ a \in {\cal G}, \qquad K(g)a=gag^{-1} \qquad \forall g \in \gru.
\eqno(\nr.1)$$
We can now introduce the concept of coadjoint orbit of $G$ (K-orbit). For
a fixed element $x$ in $\algs,$ K-orbit of $x$ is the set
$$ \olx= \{ y \in \algs \mid y=g x g^{-1} \}. \eqno(\nr.2)$$
Let us suppose that there exists a subset \grh of $\gru,$ such that
$$ h x h^{-1} = x \qquad \forall h \in H. \eqno(\nr.3)$$
One can easily prove that this set is a subgroup of $\gru,$ that we call
stability
subgroup of the point $x$. So the orbit \olx of the point $x$ is isomorphic
to the coset space $\gsh=\gru/\grh.$ \par
The properties of this construction, that are relevant for our investigations,
are listed below.\par
a) All the K-orbits of $\algs,$ i.e. the orbits of all elements of $\algs,$
are homogeneous G-spaces. Namely, the action of the group $G$ by means of
$K(g)$ maps a point of a fixed orbit $\olx$ into another point of the same
orbit.\par
b) Two different K-orbits are disjoint, i.e. we cannot  connect a point of an
orbit with another point of another distinct orbit by means of a coadjoint
representation.\par
c) All the K-orbits are symplectic manifold of even dimension\cfr{37}. One
can, in fact,
define on every orbit an exterior differential 2-form $\bbf$ which is closed,
skewsymmetric, nondegenerate. To this purpose, let us consider the mapping
$$ \partial_x \longleftrightarrow x \eqno(\nr.4)$$
that induces a correspondence between the tangent space to $\olx$ and some
subspace of
$\algs.$ Then for every point $f$ on $\olx$ we can define a differential 2-form
through the relation
$$ \bbf(x,y)=f([x,y]), \ \ \forall x,y \in \alg \ \, \ \ \forall f \in \algs.
\eqno(\nr.5)$$
This form is obviously skewsymmetric, closed because of Jacobi identity,
nondegenerate on the
factor space $\alg/\algh,$ where \algh is the Lie algebra of the
stabilizer of $f$, isomorphic to the tangent space to $\olx$ at point
$f$.\par

Starting from the 2-form (3.5) we can then easily construct a Poisson bracket
on $\olx$ through the relation
$$ \{\varphi_1,\varphi_2 \}=(\bbf)^{{ik}} \partial_i \varphi_1 \partial_k
\varphi_2. \eqno(\nr.6)$$
For a vast class of Lie algebras, all the coadjoint orbits are K\"ahler
manifold. We can, in that case, introduce on $\olx$ a complex structure \jj
$$ \jj:\olx \rightarrow \olx \ \ , \ \ \jj^2=-\id \eqno(\nr.7)$$
and construct a metric $g$ from \bbf
$$ g(x,\jj y)=\bbf(x,y), \qquad g(\jj x , \jj y ) = g (x,y). \eqno(\nr.8)$$
By means of the relations (3.5) and (3.8) we can define on the
orbit a hermitian scalar product $(x,y)$\cfr{25},
$$ (x,y)  = g(x,y)+i \bbf(x,y). \eqno(\nr.9)$$
(3.9) allows us to obtain the dual $x^\star$ of an orbit point $x$ with
respect to the scalar product (,)
and to give a resolution of
the identity (when this is possible) through the relation
(2.2).
Furthermore, in correspondence with every orbit we can construct an irreducible
unitary representation of the dynamical group. In fact if we consider the
maximal admissible subalgebra subordinate
\footnote{$^{(a)}$}{We recall that a subalgebra $ \algn \subseteq \alg $ is
subordinate to the 1-form $f$
 if the form $\bbf$ vanishes identically on \algn. A subordinate subalgebra
is called admissible if $ {\rm codim }\algn = {1 \over 2 } {\rm dim }\olf $
and $ \algn^\perp + f \subset \olf, $ where $\algn^\perp$ is
the set of elements of $\algs$, whose extensions over $\alg_c$ vanish
identically on \algn. }
to any point of the orbit we can
obviously construct a 1-dimensional representation of the corresponding
subgroup of $\gru.$ By means of Mackey induction\cfr{40,41}, from this
representation we
can recover a representation of $\gru.$ It has been shown by
Kirillov\cfr{36,37}, Kostant and
Auslander\cfr{42,43}, that for a vast class of Lie groups this is the way to
obtain all irreducible unitary representations.
In particular, for the class of the nilpotent Lie algebras,  one can
show, in a fairly straightforward way, that all coadjoint orbits that are
linear
manifolds, are coherent states manifolds for the dynamical Lie algebra
as well\cfr{38}.
It is expected that it should be possible to classify
all coherent states manifolds for a wider class of dynamical algebras
by means of the coadjoint orbits method.
\medskip
{\bf 3.2 A possible extension to the superalgebraic case.}
\smallskip
We look now for a way of generalizing the orbits method to the super case,
trying
to obtain
an algorithmic procedure which gives us all possible supersymplectic (K\"ahler)
mechanics and all possible UIR related to our dynamical superalgebra.\par
The first step is to introduce a correct generalization of the
concept of space of exterior 1-forms on the algebra. \par
Let $\alg=\algab$ be a finite dimensional Lie superalgebra
(typically, the dynamical
algebra for some physical model) with $\alga$ $(\algb)$ even (odd) part.
Let us consider in
$\alg$ a standard homogeneous base $(q_i,r_j)$ with $q_i \ (r_j) $ even (odd)
generators.  We can then define the space $\algs$ of 1-forms on $\alg,$ which
is
isomorphic to $\alg.$ $\algs$ is a \ZB-graded space, with
the standard homogeneous base $ (e_i,f_j) $ defined through the relations
$$\eqalign{e_i(q_l)= & \delta_{il}, \ \ e_i(r_k)=0 \cr
f_j(q_l)= & 0, \ \ f_j(r_k)=\delta_{jk} \cr} \eqno(\nr.10)$$
Let $\grass=\graab$ be a finite or infinite dimensional Grassmann algebra,
$\graa$ $(\grab)$ being
the even (odd) sector of $\grass.$ We introduce the
left $\graa-$module on $\alg,$ as the set of objects of the form
$$ a_i q_i + \alpha_j r_j, \ \ a_i \in \graa, \ \alpha_j \in \grab.
\eqno(\nr.11)$$
We call space of exterior even 1-forms on $\alg,$ denoted $\algls,$
the set of elements
$$ b_i e_i + \beta_j f_j, \ \ b_i \in \graa, \ \beta_j \in \grab.
\eqno(\nr.12) $$
We can identify $\algls$ as the super analogue of the dual of the
superspace $\algl,$ defining in $\algls$ a standard homogeneous base $(E_i,
F_j) $
by means of
$$\eqalign{E_l ( a_i q_i + \alpha_j r_j ) = &
a_i e_l (q_i) + \alpha_j e_l ( r_j) = a_l \cr
F_k ( a_i q_i + \alpha_j r_j ) = & a_i f_k (q_i) + \alpha_j f_k ( r_j) =
\alpha_k \cr} \eqno(\nr.13)$$
and a generic $\algls$ element by
$$ b_i E_i + \beta_j F_j. \eqno(\nr.15) $$
There are good reasons to believe that $\algl$ and $\algls$ are isomorphic (see
conjecture in section 3.3 for an intuitive proof). Then we can construct on the
space $\algls\simeq\algl$ a representation $\ko$ of
supergroup $\gru$ corresponding to superalgebra $\alg,$ obtained by
exponentiating
$\algl,$ by means of
$$ \ko x = g x g^{-1}, \eqno(\nr.16)$$
and introduce the concept of $\ko$-orbit
$$ \olx= \{y \in \algs \mid y=g x g^{-1} \}, \eqno(\nr.17)$$
in a way strictly analogous to the ordinary case.\par
Obviously, orbits constructed in this way are homogeneous $\gru-$spaces,
transitive with respect to the representation $\ko$ and pairwise
disjoint. Implementing a symplectic construction in the super case is a non
trivial task. We can,
of course, define different generalization of the 2-form $\bbf.$ The method we
adopt is to consider in every point $f$ of $\olx$ the 2-form
$$ \ccf= f([x,y]), \ \  \forall x,y \in \algls. \eqno(\nr.18)$$
The form $\ccf$ is super skewsymmetric. Since $\algls$ is a (formal) ordinary
Lie
algebra, the closure of the form $\ccf$ is guaranteed. In all examples $\ccf$
turns out to be
nondegenerate as well. Then $\ccf$ is supersymplectic.  From $\ccf$ we can
recover superpoisson brackets by means of\cfr{\,17}
$$ \def\derpar#1{\smash{\mathop{\partial}\limits^{#1}}}
\{\varphi_1,\varphi_2\}= \varphi_1 \derpar{\leftarrow}_{i} \ (\ccf)^{ik} \
\derpar{\rightarrow}_{k} \varphi_2
\eqno(\nr.19)$$
and, when the supermanifold is K\"ahler, a metric $g$ in the customary way.
\par
We are finally able to identify, among all $\ko$-orbits of a generic
contragradient Lie
superalgebra, the class of orbits which are isomorphic to the coherent
states
supermanifolds corresponding to the non degenerate representations of the same
Lie superalgebra. This class of orbits is simply obtained acting by the
$\ko$-representation on elements of the Cartan subalgebra of the
Lie superalgebra considered. Such orbits, isomorphic to coherent states
supermanifolds are of the form
$$ \olh=\{y \in \algl \mid  y=ghg^{-1} \} \ \ \ h \in \algh. \eqno(\nr.20)$$
This shows how, for non degenerate representations of contragradient
Lie superalgebras, results
obtained by means of generalized
Rasetti-Perelomov method and coadjoint orbits method are in agreement. \par
\eject
{\bf 3.3 Matrix formulation.}
\smallskip
A (faithful) linear representation of a finite dimensional Lie superalgebra
$\alg$ is an isomorphism, which maps $\alg$ into a finite dimensional matrix
Lie
superalgebra. Then every element $a \in \alg$ is a matrix. Let the
matrix Lie superalgebra have dimension $(m,n)$ over
the field \KK. Every element $a$ is a block matrix with elements in \KK, of the
form
$$ a =
\left(\matrix{
A_{11}
&
A_{12}
\cr
A_{21}
&
A_{22}
\cr
}\right),
\eqno(\nr.21)$$
with dim$A_{11}=m \times m,$ dim$A_{12}=m \times n,$ dim $A_{21}=n \times m$,
dim$A_{22}=n \times n.$ We call ${\rm Mat(m|n)}$ the set of matrices of this
kind.\par
A generic element $\alpha$ of the $\graa-$module on $\alg,$ denoted $\algl,$
has
the
same form, but matrices $A_{11}$ and $A_{22}$ have elements in \graa, while
$A_{11}$ and $A_{22}$ have elements in \grab. We denote this set of matrices by
${\rm Mat(m|n)}_\Lambda.$\par
We can construct a matrix representation of the space of even 1-forms on
$\alg,$ by means of the mapping $\langle , \rangle$ defined through the
relation
$$\eqalign{ & \langle , \rangle:
{\rm Mat(m|n)}_\Lambda \times \alg \longrightarrow \graa \cr
& \langle M_f,a\rangle ={\rm str}(M_f \cdot a ) \cr}. \eqno(\nr.22) $$
The matrix $M_f$ corresponds to the even 1-form $f, \ f(a)= \langle
M_f,a\rangle$.
We conjecture that, at least for $m \ne n,$ such
correspondence is one-to-one in general.\par
\smallskip
\conj
\item{} For all finite dimensional matrix Lie superalgebras,
there exists a one to one correspondence between matrices of even 1-forms and
elements
of the superalgebra. This correspondence is defined by the mapping
$\langle , \rangle$.\par
\smallskip
A plausibility argument for this conjecture is the following.
A finite dimensional matrix Lie superalgebra is constructed by means of
some commutation invariant relations on the matrix. We can impose the same
relations on
the set of matrices corresponding to all even 1-forms, which we construct
through the mapping $\langle , \rangle$. Thus we can attribute
this set, which is  in one to one
correspondence with even
1-forms space, the same superalgebra structure. Also for every
superalgebra element there exists an even 1-form. \par

Now, we can construct $\ko$-orbits for each superalgebra in this set,
acting with
$\ko$ on the (faithful) matrix representation of the Lie superalgebra with the
corresponding (faithful) matrix representation of the Lie supergroup, which we
obtain from superalgebra by means of the exponential mapping. Classifying
$\ko$-orbits is thus the same as classifying classes of conjugate elements of
the Lie superalgebra.\par
In order to classify $\ko$-orbits, we classify first all subsuperalgebras of
the
superalgebra, and then we consider only those whose corresponding
subsupergroup stabilizes a point of the space of the even 1-forms, i.e. of the
superalgebra.\par
We finally construct the generalized Kirillov-Kostant 2-form in the same
manner as in (3.18),
where superalgebra elements are now matrices.\par
In the following section we describe a simple  application as
exemplification of the method just exposed.
\medskip
{\bf 3.4 Weyl-Heisenberg superalgebra.}
\smallskip
Let us consider the five dimensional Lie superalgebra $\algw,$ with the set of
generators
$$ \{ \id , \a ,\as, \b , \bs \} $$
and basic commutation relations
$$\eqalign{[ \a, \as ] = & \{ \b , \bs \} = \id \cr
[ \a , \id ] = [ \b , \id ] = & [\a,\b]=[\as,\b]=0. \cr}$$
We can introduce a new set of generators
$$ e_1= { 1 \over \sqrt{2}}(\as + \a ) \ \ , \ \
   e_2= { i \over \sqrt{2}}(\as - \a ) \ \ , \ \
   e_3= i \id \ \ , \ \
   e_4= \b \ \ , \ \
   e_5= i \bs.
$$
In this basis, a faithful five-dimensional matrix representation for a generic
element in
$\algwl$ is
$$
x
=
\left(\matrix{
0
&
x_1
&
x_3
&
\alpha_1
&
\alpha_2
\cr
0&0&
x_2
&0&0
\cr
0&0&0&0&0
\cr
0&0
&
\alpha_2
&
0
&
0
\cr
0&0
&
\alpha_1
&0&0
\cr
} \right) \ \ \ \ x_i \in \rr_c, \ \  \alpha_j \in \rr_a.$$
The matrix $M_f$ of generic exterior even 1-form $f$ on $\algw$ takes the form
$$
M_f=\left(\matrix{
0&0&0&0&0
\cr
y_1&0&0&0&0
\cr
y_3&y_2&0&0&0
\cr
\delta_1&0&0&0&0
\cr
\delta_2&0&0&0&0
\cr
} \right) \ \ \ \ y_i \in \rr_c, \ \  \delta_j \in \rr_a.$$
Exponentiating the generic element $w$ of $\algwl,$ we can obtain a generic
element $g$ of the corresponding supergroup $\grw.$ By utilizing Berezin's
decomposition $g=g_e \circ g_o$, we recover $g_e \ (g_o)$ by
exponentiating the even (odd) part of $w$.
$$
g=g_e \circ g_o =
\left(\matrix{
1&x_1&x_3+{1 \over 2}x_1 x_3&0&0
\cr
0&1&x_2&0&0
\cr
0&0&1&0&0
\cr
0&0&0&1&0
\cr
0&0&0&0&1
\cr
} \right)
\cdot
\left(\matrix{
1&0&0&\alpha_1&\alpha_2
\cr
0&1&0&0&0
\cr
0&0&1&0&0
\cr
0&0&\alpha_2&1&0
\cr
0&0&\alpha_1&0&1
\cr
} \right) $$
whose inverse is
$$ g^{-1}=
\left(\matrix{
1&-x_1&-x_3+{1 \over 2}x_1x_2&-\alpha_1&-\alpha_2
\cr
0&1&-x_2&0&0
\cr
0&0&1&0&0
\cr
0&0&-\alpha_2&1&0
\cr
0&0&-\alpha_1&0&1
\cr
} \right).
$$
The $\ko$-orbit in generic position has thus matrix form
$$ \olf= P \{ g_o^{-1} g_e^{-1} f g_e g_o \} =
\left(\matrix{
0&0&0&0&0
\cr
y_1+x_2 y_3&0&0&0&0
\cr
y_3&y_2-x_1y_3&0&0&0
\cr
\delta_1 + \alpha_2 y_3&0&0&0&0
\cr
\delta_2+ \alpha_1 y_3 &0&0&0&0
\cr
} \right),
$$
where $P$ is the projection from ${\rm Mat(3|2)}_\Lambda$ onto
$\algwl^\star$.
$\ko$-representation maps coordinates of the point $f$ into points
$$
\eqalign{y_1 & \longrightarrow y_1 + x_2 y_3\cr
y_2 & \longrightarrow y_2 - x_1 y_3\cr
y_3 & \longrightarrow y_3       \cr
\delta_1 & \longrightarrow \delta_1 + \alpha_2 y_3\cr
\delta_2 & \longrightarrow \delta_2 + \alpha_1 y_3. \cr}
$$
There are obviously only two classes of orbits, in analogy with the ordinary
Weyl-Heisenberg algebra case: \par
a) Orbits of the points $ f_0=\{ y_1,y_2,0\mid \delta_1,\delta_2 \}$
that are points in
$\algws$ labelled by their coordinate set. The corresponding supermanifold is
trivial. The representations
$T^{y_1,y_2,\delta_1,\delta_2}$ constructed from orbits
belonging to this class are 1-dimensional UIR. In fact, the maximal admissible
subsuperalgebra subordinate to the 1-form $f_0$ is the whole superalgebra
$\algw.$ We obtain
$$ T^{y_1,y_2,\delta_1,\delta_2}= e^{i(y_1 x_1+y_2x_2+\delta_1 \alpha_1
+\delta_2 \alpha_2)}. $$
\par
b) Orbits of the points $f_1$, which have $ y_3 \ne 0 $ are four dimensional
superplanes labelled by the value of this coordinate. On this class of orbits,
the matrix of the generalized supersymplectic 2-form $\ccf$ in the
$ \{ x_1,x_2,\alpha_1,\alpha_2 \}$ basis is
$$
\ccf =y_3 \cdot
\left(\matrix{
0&1&0&0
\cr
-1&0&0&0
\cr
0&0&0&1
\cr
0&0&1&0
\cr
} \right).
$$
$\ccf$ is closed and nondegenerate. We can define on the orbit a complex
structure $\jj$ of type (3.7) by means of the relations
$$\eqalign{ & \jj:x_1 \mapsto -x_2, \ \ \jj:x_2 \mapsto x_1 \cr
& \jj:\alpha_1 \mapsto -\alpha_2, \ \ \jj:\alpha_2 \mapsto \alpha_1. \cr} $$
{}From the latter we can construct on the orbit a metric tensor as in (3.8),
by writing first the K\"ahler potential
$$
K=y_3 (x_1x_2 + \alpha_1 \alpha_2)
$$
whereby the metric is recovered by the relation
$$ g_{ij}= {\partial K \over \partial z_i \partial z_j }, $$
with $ z_i,z_j $ in $\{x_1,x_2,\alpha_1,\alpha_2 \}$.
The superpoisson brackets, constructed resorting to (3.19), are
$$\def\derpar#1{\smash{\mathop{\partial}\limits^{#1}}}
\{\varphi_1,\varphi_2\}_s= \partial_a \varphi_1 \partial_b \varphi_2 -
\partial_b \varphi_1 \partial_a \varphi_2 +
\varphi_1 \derpar{\leftarrow}_\alpha  \derpar{\rightarrow}_\beta \varphi_2 +
\varphi_1 \derpar{\leftarrow}_\beta  \derpar{\rightarrow}_\alpha \varphi_2. $$
The UIR's of supergroup $\grw$ corresponding to this class of orbits, can be
induced from the 1-dimensional representations
$$ U^{y_3}(x_3)= e^{iy_3 x_3} $$
of the 1-dimensional subsupergroup of \grw, generated by the center of Lie
superalgebra $\algw,$ which is the maximal admissible subalgebra subordinate
to the 1-form $f_1.$\par
These results are in concordance with those obtained in the ordinary way by
Nieto\cfr{13}.\par

\bigskip
\centerline{\bf 4. EXAMPLES}
\medskip
\bf 4.1 The simplest case: $u(1|1)$ \par
\smallskip
\rm
$l(1|1)$ is the superalgebra of $(1+1) \times (1+1)$ matrices, with
$$C = \pmatrix{1 & 0 \cr 0 & 1 \cr}, \qquad H = \pmatrix{1 & 0 \cr 0 & 0 \cr}$$
as even generators,
$$E_+ = \pmatrix{0 & 1 \cr 0 & 0 \cr}, \qquad
E_- = \pmatrix{0 & 0 \cr 1 & 0 \cr}$$
as odd generators, and commutation-anticommutation relations
$$[H,E_\pm] = \pm E_\pm \qquad \{ E_+,E_- \} = C; $$
all other commutators-anticommutators being equal to zero. \par
$l(1|1)$ can be viewed as the contragradient Lie superalgebra of rank 1
and null
Cartan matrix, extended by the Cartan element $H$, and
can be realized in terms of fermionic creation-annihilation operators, by
means of the correspondence
$$C \mapsto \ii, \qquad H \mapsto a^\dagger a,
\qquad E_+ \mapsto a^\dagger, \qquad E_- \mapsto a. $$
\eject
\underbar{Coherent states supermanifold} \par
The defining representation is equivalent to that in one-fermion Fock space,
and has as its highest weight vector
$$|\psi_0\rangle = \pmatrix{1 \cr 0 \cr}.$$
$u(1|1)$ is the compact form of $l(1|1)$ defined by
$$u(1|1) = \left\{ M = i\phi_0 C + i\phi H + \eta E_+ - \eta^* E_- |
\phi_0, \phi \in \rr_c, \eta \in \Lambda_1 \right\}, $$
where $\rr_c = \{ \lambda \in \Lambda_0 | \lambda^* = \lambda \}$.
The exponential group of matrices associated with $u(1|1)$, denoted by
$U(1|1)$, is the group of matrices
$$g = \pmatrix{a & \beta \cr \gamma & d \cr}; \qquad a,d \in \Lambda_0,
\, \beta, \gamma \in \Lambda_1 $$
such that $g^+g = gg^+ = \ii$, where the adjoint is defined by
$$g^+ = \pmatrix{a^* & \gamma^* \cr \beta^* & d^* \cr}. $$
Group elements can be written in the form $g = \xi \cdot u$, with $\xi =
\exp (\eta E_+ - \eta^* E_-)$ and $u = \exp (i\phi_0 C + i\phi H)$. The
isotropy subgroup is just $U(1) \otimes U(1)$, generated by the $u$'s,
and the SCS supermanifold is the coset space $\displaystyle
\mm = {U(1|1) \over U(1) \otimes U(1)}$, parametrized by the
coset representatives $x \equiv x(\eta, \eta^*) = \xi$. \par
We are thus able to apply the method described at the end of previous
chapter: acting on the point $x \in \mm$ by the infinitesimal group element
$$\delta g = \exp[dt(ig^0 C + ig^1 H + \zeta E_+ - \zeta^* E_-)], $$
we obtain the variation of the coordinates on the supermanifold:
$${d\eta \over dt} = ig^1\eta + \zeta, \qquad
{d\eta^* \over dt} = -ig^1\eta^* + \zeta^*. $$
The Cartan-Killing form on $l(1|1)$ is degenerate, so we define the
nondegenerate form
$\gamma_d$ by
$$\gamma_d(X,Y) = {\rm str}(XY) \qquad \forall X,Y \in l(1|1).$$
Finally, (2.15) gives
$g^{\eta \eta^*} = - g^{\eta^* \eta} = 1$, from which we obtain the metric
$$g_{\eta^* \eta} = - g_{\eta \eta^*} = 1, \qquad
g = \pmatrix{0 & -1 \cr 1 & 0 \cr}.$$
This admits the natural complex structure defined by
$$\jj\left({\displaystyle{\partial \over \partial \eta}}\right) = i
{\displaystyle{\partial \over \partial \eta}}, \qquad
\jj\left({\displaystyle{\partial \over \partial \eta^*}}\right) = -i
{\displaystyle{\partial \over \partial \eta^*}}$$ that one can use to
define the 2-form $\omega = -2i d\eta \wedge d\eta^*$, which is closed. Then
$g$ is K\"ahler, with
K\"ahler potential $K(\eta,\eta^*) = \eta \eta^*$, and induces the invariant
measure $d\mu(x) \equiv 1$, as we could expect from the invariant measure
$d\mu(g) \equiv 1$ on the $U(1|1)$ group manifold. \par
\smallskip
\underbar{Coadjoint orbits approach} \par
Let us consider the superalgebra \uu. A generic element of the left
$\graa-$module over \uu, that we denote \uul, takes the matrix form
$$ e = \left(\matrix{ia&\delta\cr-\delta^*&i d\cr}\right)  \ \ \ \ a,d
\in \rr_c, \ \ \delta \in \grab. $$
Even though $m = n$, we can resort here to the conjecture exposed in
section 3.3 giving to the space $\uuls$ of exterior even 1-forms on
$\uul$ the form
$$ M_f = \left(\matrix{ix&\alpha\cr-\alpha^*&i y\cr}\right) \ \ \ \ x,y \in
\rr_c, \ \ \alpha \in \grab.$$
The orbit $\olf$ of $f$, obtained from (3.17) takes then the matrix form
$$ \olf= \left(\matrix{
i (1-\delta \delta^* ) x- \alpha \delta^* \exp (-i u)-
&
i \delta (x-y) + \alpha \exp (-iu)
 \cr
\ \ \ \ \ - \alpha^* \delta \exp(iu) +i \delta \delta^* y
&
\
\cr
\ & \ \cr
+i \delta^*(x-y)-\alpha^* \exp(iu)
&
i (1+\delta \delta^*
)y- \alpha \delta^* \exp (-iu)
\cr
\
&
\ \ \ \ \ - \alpha^* \delta \exp(iu)  -i \delta \delta^* x
\cr}\right) $$
where $u=a-d$.\par
We can classify $u(1|1)_\Lambda$'s $\ko$-orbits, by means of the
method of subsuperalgebras
exposed in section 3.4. As we conjectured above
(section 3.3), since $\uu$ is contragradient, orbits of elements of the kind
$$ f_1 = \left(\matrix{i x&0\cr0&i y \cr}\right) \ \ \ \ x \ne y \land x \ne
0 \land y \ne 0 $$
are isomorphic to coherent states supermanifolds corresponding to nondegenerate
representations of \uul$.$ The subgroup that stabilizes points belonging
to this class is the (ordinary) subgroup $ U(1) \otimes U(1) $. The orbits are
supermanifolds of dimension $(0,2)$. The generalized Kirillov-Kostant 2-form
\ccf$,$ which is nondegenerate, has matrix form
$$ \ccf = i(x-y) \left(\matrix{0&1\cr1&0\cr}\right). $$
We can introduce a complex structure of kind (3.7) by means of the relations
$$ \jj:Re(\delta) \longmapsto -Im(\delta), \ \ \jj: Im(\delta) \longmapsto
Re(\delta) $$
and recover a metric tensor of the form (3.8), which admits a K\"ahler
potential
$$ K= (x-y)\delta \delta^*. $$
The superpoisson brackets, defined by (3.19), can be finally written as
$$ \def\derpar#1{\smash{\mathop{\partial}\limits^{#1}}}
\{\varphi_1,\varphi_2\}=
\varphi_1 \derpar{\leftarrow}_{\!\delta}\derpar{\rightarrow}_{\!\delta^*}
\varphi_2
+ \varphi_1 \derpar{\leftarrow}_{\!\delta^*}\derpar{\rightarrow}_{\!\delta}
\varphi_2
$$
There are four other classes of orbits, which, however, are not relevant for
our
scope.\par
\medskip
\bf 4.2 Another "ordinary unitary" example: $su(2|1)$ \par
\smallskip
\rm
$sl(2|1)^{44}$ is the superalgebra of $(2 + 1) \times (2 + 1)$ supermatrices
with zero supertrace. It has even sector generated by $\{Q,J_3,J_+,J_-\}$
and odd sector generated by $\{ W_+,W_-,V_+,V_-\}.$\par
These generators satisfy the following commutation rules:
$$\eqalign{& [J_3,J_\pm]=\pm J_\pm \qquad [J_+,J_-]=2J_3 \cr
& [Q,J_3]=[Q,J_\pm]=0 \cr
& [Q,V_\pm]={1 \over 2}V_\pm \qquad [Q,W_\pm]=-{1 \over 2}W_\pm \cr
& [J_3,V_\pm]=\pm{1 \over 2}V_\pm \qquad [J_3,W_\pm]=\pm{1 \over 2}W_\pm \cr
& [J_\pm,V_\mp]=V_\pm \qquad [J_\pm,W_\mp]=W_\pm \cr
& [J_\pm,V_\pm]=[J_\pm,W_\pm]=0 \cr
& \{V_\pm,V_\pm\}=\{V_\pm,V_\mp\}=0 \cr
& \{W_\pm,W_\pm\}=\{W_\pm,W_\mp\}=0 \cr
& \{V_\pm,W_\pm\}=\pm J_\pm \qquad \{V_\pm,W_\mp\}=-J_3\pm Q, \cr}$$
from which one can easily realize that the $J$'s are the generators of
a Lie algebra $su(2)$, whereas $Q$ generates a commuting $u(1)$. \par
Before defining $su(2|1)$ (of which $sl(2|1)$ is the complexification)
we must summarize a few properties of the irreducible representations (IRREP)
of
$sl(2|1)^{44}$. \par
These IRREP are labelled by two quantum numbers, $j$ and $q$
(referred to as spin and charge, respectively), which are defined by
$$J^2|\psi_0\rangle=j(j+1)|\psi_0\rangle, \qquad
J_3|\psi_0\rangle=j|\psi_0\rangle, \qquad
Q|\psi_0\rangle=q|\psi_0\rangle, $$
$|\psi_0\rangle$ denoting the highest weight vector,
$j \in \{ 0, {1 \over 2}, 1, \ldots \}$ and $q \in \cc$. The
representation space may contain up to 4 multiplets spanned by the
$(Q, J^2, J_3)$ eigenvectors
$$\eqalign{& |q,j,m\rangle, \qquad m=-j,-j+1,\ldots,j \cr
& |q \pm {\textstyle{1 \over 2}},j-{\textstyle{1 \over 2}},m\rangle, \qquad
m=-j+{\textstyle{1 \over 2}},-j+{\textstyle {3 \over 2}},\ldots,
j-{\textstyle{1 \over 2}} \cr
& |q,j-1,m\rangle, \qquad m=-j+1,-j+2,\ldots,j-1. \cr}$$
Apart from the trivial one-dimensional one, there are three series of
IRREP: \par
a) a nondegenerate series, in which the highest weight vector is
annihilated only by the raising operators $J_+, V_+, W_+$; \par
b) a degenerate series, in which $W_-$ also annihilates $|\psi_0\rangle$; \par
c) a second degenerate series, obtained from the previous one by means of the
automorphism
$$J_3 \mapsto J_3, \qquad J_\pm \mapsto J_\pm, \qquad Q \mapsto -Q, \qquad
V_\pm \mapsto W_\pm, \qquad W_\pm \mapsto V_\pm. $$
Finally, we can introduce in $sl(2|1)$ the "normal adjoint"
operation $^\dagger$ defined by
$$Q^\dagger = Q, \qquad J_3^\dagger = J_3, \qquad
V_\pm^\dagger = \pm W_\mp, \qquad W_\pm^\dagger = \mp V_\mp. $$
It is the possibility of defining a "normal adjoint" (that is, an involution
with
all the properties of Hermitian adjoint) that justifies the different
approaches of present section and of the next one. \par
Turning now to the superalgebra $su(2|1)$, it is but the compact form of
$sl(2|1)$ defined by
$$\eqalign{su(2|1)=\{ & i\phi_0Q+i\phi J_3+zJ_+-z^*J_-+\eta_1W_++\eta_1^*V_-
+\eta_2W_--\eta_2^*V_+| \cr
& \phi_0,\phi\in \rr_c, z \in \Lambda_0, \eta_k \in \Lambda_1\}. \cr} $$
The exponential group associated with $su(2|1)$, denoted by $SU(2|1)$,
is the group of $(2 + 1) \times (2 + 1)$ unitary, unimodular supermatrices.
\par \smallskip
\underbar{Coherent states supermanifold} \par
Looking at the IRREP's theory of $sl(2|1)$ one can understand how
SCS (and their supermanifold) depend on the choice of the IRREP of
$sl(2|1)$ from which the UIR of $SU(2|1)$ is obtained. Taking an IRREP of
the (a) series, the isotropy subgroup of $|\psi_0\rangle$ will be
$U(1) \otimes U(1)$, generated by $Q$ and $J_3$. Taking an IRREP of the (b)
series, the isotropy subgroup will be a true supergroup, $U(1|1)$,
generated by $Q, J_3, V_+, W_-$. \par
We study in detail the first
case, constructing first an $SU(2|1)$-invariant metric on the SCS
supermanifold $\mm = {\displaystyle{SU(2|1) \over U(1) \otimes U(1)}}$
and trying to establish whether $\mm$ is K\"ahler or not.  \par
The group elements can again be written in the form $g = \xi \cdot u$,
where now $\xi=\exp(\eta_1W_++\eta_1^*V_-+\eta_2W_--\eta_2^*V_+)$ and
$u=\exp(i\phi_0Q+i\phi J_3+zJ_+-z^*J_-)$, and then the coset
representatives are of the form $x = \xi \cdot v$, with
$v=\exp(zJ_+-z^*J_-)$. \par
Acting on the point $x \in \mm$ by the infinitesimal group element
$$\delta g= \ii + \left(
\sum_{i=0}^3 g^iA_i+\zeta^1W_++{\zeta^1}^*V_-+\zeta^2W_--{\zeta^2}^*V_+
\right) dt, $$
where $A_0 = iQ, A_1 = {\displaystyle{i \over 2}}(J_+ + J_-),
A_2 = {\displaystyle{1 \over 2}}(J_- - J_+), A_3 = iJ_3$,
after the substitutions
$-i {\displaystyle{z \over |z|}} \tan|z| \mapsto z$ and
$${\tanh\sqrt{-BB^\dagger} \over \sqrt{-BB^\dagger}}B \mapsto
\pmatrix{\eta_1 \cr \eta_2 \cr}, \qquad B = \pmatrix{\eta_1 \cr \eta_2 \cr}$$
we obtain (introducing the simplified notation $e_i = \eta_i^* \eta_i$)
the variations of the coordinates on the supermanifold
$$\eqalign{{dz \over dt}=
& ig^3z+{1 \over 2}g^1(1+z^2)+{i \over 2}g^2(1-z^2) \cr
& -{1 \over 2}\left[({\zeta^1}^*\eta_1+\zeta^1\eta_1^*)
\left(1-{1 \over 2}e_2\right)-({\zeta^2}^*\eta_2+\zeta^2\eta_2^*)
\left(1-{1 \over 2}e_1\right)\right]z \cr
& +{i \over 2}\left[(\zeta^1\eta_2^*+z^2{\zeta^1}^*\eta_2)
\left(1-{1 \over 2}e_1\right)+
({\zeta^2}^*\eta_1+z^2\zeta^2\eta_1^*)
\left(1-{1 \over 2}e_2\right)\right], \cr
{d\eta_1 \over dt}= &
-{i \over 2}(g^0-g^3)\eta_1+{i \over 2}(g^1+ig^2)\eta_2+\zeta^1+
{\zeta^2}^*\eta_2\eta_1, \cr
{d\eta_2 \over dt}= &
-{i \over 2}(g^0+g^3)\eta_2+{i \over 2}(g^1-ig^2)\eta_1+\zeta^2+
{\zeta^1}^*\eta_1\eta_2. \cr} \eqno(4.1)$$
The Cartan-Killing form is nondegenerate
and thus from (2.15) we obtain the inverse of the metric
tensor, which has the form $g^{-1}=\pmatrix{A & B \cr C & D \cr}$
(the explicit expression of each block is given in the Appendix).
The metric can be obtained by taking the inverse of this block
supermatrix, and is straightforwardly given by the formula
$$g =
\pmatrix{\ii & 0 \cr -D^{-1}C & \ii \cr} \cdot
\pmatrix{\left(A-BD^{-1}C\right)^{-1} & 0 \cr 0 & D^{-1} \cr} \cdot
\pmatrix{\ii & -BD^{-1} \cr 0 & \ii \cr}. $$
{}From the latter (we do not report it explicitly because too cumbersome) we
can
calculate the invariant measure
$$d\mu(x)=\left({\rm sdet}(g^{-1})\right)^{-1/2}=(1+z^*z)^{-2}(1+e_1+e_2),
$$
which allows us to derive information relevant to our main question. \par
The SCS supermanifold $\mm$ can be written (see Sec. 2.2) as the product
$\displaystyle {SU(2|1) \over U(2)}
\textstyle{\otimes_s} \displaystyle{U(2) \over U(1) \otimes
U(1)}$, where the second (even) factor is the 2-dimensional
sphere $S_2 \equiv \cc P^1$ | an Einstein-K\"ahler symmetric space.
The first (odd) factor is a symmetric space in some sense similar to $\cc
P^2 \equiv \displaystyle {SU(3) \over U(2)}$. This is
again a K\"ahler supermanifold, with invariant metric
$$g = D^{-1} = \pmatrix{
0 & 1-\eta_2^*\eta_2 & 0 & \eta_1^*\eta_2 \cr
-(1-\eta_2^*\eta_2) & 0 & -\eta_2^*\eta_1 & 0 \cr
0 & \eta_2^*\eta_1 & 0 & 1-\eta_1^*\eta_1\cr
-\eta_1^*\eta_2 & 0 & -(1-\eta_1^*\eta_1) & 0 \cr}, $$
and K\"ahler potential (up to an additive
constant) given by
$K=\eta_1^*\eta_1+\eta_2^*\eta_2-\eta_1^*\eta_1\eta_2^*\eta_2$.
It is worth pointing out that this K\"ahler potential can be obtained by taking
the
logarithm of the invariant measure function
$$d\mu(x)=(1+\eta_1^*\eta_1+\eta_2^*\eta_2),
$$
exactly as for an Einstein space. \par
We have thus proved that our supermanifold is the (semidirect) product of two
Einstein-K\"ahler symmetric spaces. We conjecture that it is K\"ahler as well;
but
not Einstein. In fact, if it were, its K\"ahler potential would be the sum
of the K\"ahler potentials of the two factor manifolds, and the metric
would have a block matrix form, thus exhibiting a complete separation
(orthogonality) between even and odd coordinates. However, this is not the
case,
in that it appears evident from (4.1) that the variation of even coordinates
under the group action depends on the odd ones. \par
\smallskip
\underbar{Coadjoint orbits approach} \par
Let us consider the contragradient Lie superalgebra \su. The
(fundamental) matrix representation of a generic $\sul$ element is
$$ e =
\left(\matrix{
i(a+b)&
c&
\alpha\cr
-c^*&
i(a-b)&
\beta\cr
-\alpha^*&
-\beta^* &
2 i a \cr} \right ) \ \ \ \  a,b,c \in \rr_c, \ \ \alpha,\beta \in \grab.
$$
Because of the conjecture formulated in section 3.3, we are able to equip the
space $\suls$ of even 1-forms on $\sul$ with the same superalgebra structure
as for \sul.
Then, the matrix $M_f$ corresponding to a generic 1-form $f$ takes the form
$$ M_f =
\left(\matrix{
i(x+y)&z&\xi\cr
-z^*&i(x-y)&\chi\cr
-\xi^*&-\chi^* &2 i x \cr
} \right )
\ \ \ \  x,y \in \rr_c, \ \ z \in \graa, \ \ \chi,\xi \in \grab.
$$
As shown before, we can recover representation $\ko$ by exponentiating the
fundamental representation. Thus we are able to study a generic
$\ko$-orbit. We restrict here our attention to some cases interesting for
the physical applications we have in mind.
Applying the conjecture exposed in section 3.4, the
orbits isomorphic to the coherent states supermanifolds are those obtained
acting trough $\ko$ on elements of the form
$$ f_0 =
\left(\matrix{
i(x+y)&0&0\cr
0&i(x-y)&0\cr
0&0&2 i x \cr
} \right ) \ \ \ \ x \ne y \land x \ne 0 \land y \ne 0.$$
The elements belonging to this class are stabilized by the (ordinary)
Lie subgroup $U(1) \otimes U(1)$. Also, orbits of this form are isomorphic to
the coset space $\displaystyle{SU(2|1) \over U(1) \otimes U(1)}$.
We can obtain another interesting class of orbits starting from elements of
$\suls$ of type
$$ f_1 =
ix \left(\matrix{
1
&
0
&
0
\cr
0
&
1
&
0
\cr
0
&
0
&
2
\cr} \right ) \ \ \ \ x \ne 0.$$
These are elements stabilized by the (ordinary) Lie subgroup
$U(2)=SU(2) \otimes U(1)$.
Orbits that belong to this class are isomorphic to the coset space
$\displaystyle{SU(2|1) \over U(2)}$,
which is a supermanifold of dimension (0,4).
In the Appendix we report the generalized Kirillov-Kostant 2-form. Some of its
properties are not yet known. \par

\medskip
\bf 4.3 The "graded unitary" example: $uosp(1|2)$ \par
\smallskip
\rm
So far we have used expressions like "normal adjoint", "ordinary
conjugation in a Grassmann algebra" without completely explaining what
these expressions mean. Before studying the SCS supermanifold
for $uosp(1|2)$ we must briefly discuss how hermitian conjugation in a Lie
algebra and complex conjugation in the field of complex numbers generalize
to the graded case. Let us start with hermitian conjugation. \par
Hermitian conjugation can be generalized in two different ways to the case
of Lie
superalgebras$^{34}$. The first one, called "normal adjoint" and
denoted by $^\dagger$, is defined by the ordinary axioms
$${\rm deg}(X) = {\rm deg}(X^\dagger) $$
$$(aX+bY)^\dagger=\bar a X^\dagger+\bar b Y^\dagger$$
$$[X,Y]^\dagger=[Y^\dagger,X^\dagger]$$
$$(X^\dagger)^\dagger=X,$$
where $X,Y \in {\cal G}$, $a,b \in \cc$, the bar means complex conjugation
and $[,]$ is the graded commutator. In $l(1|1)$ and $sl(2|1)$ we have
already defined operations of this kind. The second one, called "grade adjoint"
and denoted by $^\ddagger$, is defined by the following axioms
$$\eqalign{ & {\rm deg}(X) = {\rm deg}(X^\ddagger) \cr
& (aX+bY)^\ddagger=\bar a X^\ddagger+\bar b Y^\ddagger \cr
& [X,Y]^\ddagger=(-)^{XY}[Y^\ddagger,X^\ddagger] \cr
& (X^\ddagger)^\ddagger=(-)^X X, \cr}$$
where $X$ and $Y$ must be homogeneous elements and an element and its
degree are represented by the same symbol. Given a matrix
representation of a superalgebra over the complex field we have
$$M=\pmatrix{A&B\cr C&D\cr}, \qquad
M^\dagger=\pmatrix{A^+&C^+\cr B^+&D^+\cr}, \qquad
M^\ddagger=\pmatrix{A^+&-C^+\cr B^+&D^+\cr},$$
where $^+$ denotes hermitian conjugate.
Finally, the generalization of hermitian representations will be called
"star" or "grade star", depending on the kind of adjoint that has been
defined in the superalgebra. \par
A very similar situation occurs when one defines conjugation in a Grassmann
algebra$^{45}$. The ordinary conjugation, denoted by $^*$, satisfies
$$\eqalign{ & {\rm deg}(\lambda) = {\rm deg}(\lambda^*) \cr
& (c\lambda + d\mu)^* = \bar c \lambda^* + \bar d \mu^* \cr
& (\lambda\mu)^* = \mu^* \lambda^* \cr
& (\lambda^*)^* = \lambda \cr}$$
where $\lambda, \mu \in \Lambda$ and $c,d \in \cc$, while the graded
conjugation, denoted by $^\diamondsuit$, obeys
$$\eqalign{ & {\rm deg}(\lambda) = {\rm deg}(\lambda^\diamondsuit) \cr
& (c\lambda + d\mu)^\diamondsuit =
\bar c \lambda^\diamondsuit + \bar d \mu^\diamondsuit \cr
& (\lambda\mu)^* = \lambda^\diamondsuit \mu^\diamondsuit \cr
& (\lambda^\diamondsuit)^\diamondsuit = (-)^\lambda \lambda, \cr}$$
with $\lambda$ and $\mu$ homogeneous elements. \par
In our examples we deal with compact forms that have the general structure
$(\Lambda \otimes {\cal G})_0$, because even (odd) parameters are
associated to even (odd) generators; in these algebras the hermitian
conjugation $^+$ can be defined either by
$$(\lambda X + \mu Y)^+ = \lambda^* X^\dagger + \mu^* Y^\dagger $$
or by
$$(\lambda X + \mu Y)^+ =
\lambda^\diamondsuit X^\ddagger + \mu^\diamondsuit Y^\ddagger, $$
depending on which kind of adjoint can be defined in ${\cal G}$;
in both cases, $\forall Z, W \in (\Lambda \otimes {\cal G})_0$, we have
$$(Z^+)^+ = Z \qquad [Z,W]^+ = [W^+,Z^+]. $$\par
$osp(1|2)^{46}$ is the superalgebra of $(1 + 2) \times (1 + 2)$ supermatrices
with $\{J_3, J_+, J_-\}$
as even generators and $\{R_+, R_-\}$
as odd generators. The following commutations rules are valid:
$$\eqalign{ & [J_3,J_\pm]=\pm J_\pm,\qquad [J_+,J_-]=2J_3 \cr
& [J_3,R_\pm]=\pm{1 \over 2}R_\pm,\qquad [J_\mp,R_\pm]=R_\mp, \qquad
[J_\pm,R_\pm]=0 \cr
& \{R_\pm,R_\pm\}=\pm{1 \over 2}J_\pm, \qquad \{R_+,R_-\}=-{1 \over 2}J_3, \cr}
$$
from which one sees that the even sector of $osp(1|2)$ is just $su(2)$. The
IRREP of $osp(1|2)$ are labelled by the $su(2)$-quantum number $j$ and
the representation space is the direct sum of two $su(2)$ representation
spaces, one with spin $j$ and even (odd) degree and the other with spin
$j - {1 \over 2}$ and odd (even) degree. Depending on the choice of the
parity of the two subspaces (and then of the highest weight vector) one can
extend the $su(2)$ hermitian conjugation ($J_3^\dagger = J_3,
J_\pm^\dagger = J_\mp$) to a grade adjoint operation over the whole
superalgebra, by setting $R_+^\ddagger = \pm R_-$ (the upper sign
corresponding to odd highest weight vector), and consequently $R_-^\ddagger
= \mp R_+$. It is easy to verify that the defining representation has $j =
{1 \over 2}$ and odd highest weight vector. \par
The algebra $uosp(1|2)$ can now be defined by
$$\eqalign{uosp(1|2) = & \{
X=i\phi J_3 + zJ_+ - z^\diamondsuit J_- + \eta^\diamondsuit R_+ + \eta R_-| \cr
& \phi \in \rr_c, z \in \Lambda_0, \eta \in \Lambda_1 \}.\cr} $$
The exponential group associated with $uosp(1|2)$, denoted by $UOSP(1|2)$,
is the group of $(1 + 2) \times (1 + 2)$ orthosymplectic, unimodular
supermatrices. \par \smallskip
\underbar{Coherent states supermanifold} \par
The IRREP being nondegenerate, the isotropy subgroup of the highest
weight vector is just the Cartan subgroup, generated by $J_3$, the
SCS supermanifold is the coset space $\mm =
{\displaystyle{UOSP(1|2) \over U(1)}}$, and the coset representatives can
be written in the usual form $x = \xi \cdot v$, where
$\xi = \exp(\eta^\diamondsuit R_+ + \eta R_-)$ and
$v = \exp(zJ_+ - z^\diamondsuit J_-)$. Acting on the coset representative
by the infinitesimal group element $\delta g = \ii +
\left(\sum_{i = 1}^3 g^iA_i + \zeta^\diamondsuit R_++\zeta R_-\right)dt, $
after the substitution
$-i {\displaystyle{z \over |z|}} \tan |z| \mapsto z$
we obtain the variation of the coordinates
$$\eqalign{{dz \over dt} = & {1 \over 2}g^1(1 + z^2) +
{i \over 2}g^2(1 - z^2) + ig^3z \cr
& + {1 \over 4} \zeta (i\eta + z\eta^\diamondsuit) +
{1 \over 4} \zeta^\diamondsuit (z\eta - iz^2\eta^\diamondsuit), \cr
{d\eta \over dt} = & {i \over 2}(g^1+ig^2)\eta^\diamondsuit +
{i \over 2}g^3\eta + \zeta\left(1+{1 \over 4}\eta^\diamondsuit\eta\right), \cr
{d\eta^\diamondsuit \over dt} = & {i \over 2}(g^1-ig^2)\eta -
{i \over 2}g^3\eta^\diamondsuit +
\zeta^\diamondsuit\left(1+{1 \over 4}\eta^\diamondsuit\eta\right). \cr}
$$
The Cartan-Killing form is
nondegenerate and, once more,
from (2.15) we obtain the inverse of the metric tensor
and from this the
invariant measure
$$d\mu(x) = (1 + z^\diamondsuit z)^{-2}
\left( 1 - {1 \over 4} \eta^\diamondsuit \eta \right).$$ \par
Once more we can write $\mm$ as a product $\displaystyle {UOSP(1|2)
\over SU(2)}
\textstyle{\otimes_s}
\displaystyle{SU(2) \over U(1)}$, and again the bosonic factor is
$S_2$. The invariant metric on the fermionic factor is
$$g=\pmatrix{0 & 1+\displaystyle{1 \over 4}\eta^\diamondsuit\eta \cr
-\left(1+\displaystyle{1 \over 4}\eta^\diamondsuit\eta\right) &
0 \cr},$$
that admits a natural complex structure, from which the 2-form
$$\omega = 2i \left(1 - {1 \over 4} \eta^\diamondsuit \eta \right)
d\eta \wedge d\eta^\diamondsuit $$
can be derived. In this case the 2-form is
not closed. Indeed, closure and supersymmetry imply that the 2-form should
be constant, and hence manifestly non-invariant.
Then the fermionic factor is not a K\"ahler supermanifold and we assume
that the same holds for the SCS supermanifold. \par \smallskip
\underbar{Coadjoint orbits approach} \par
Let us consider the contragradient Lie superalgebra $\uo$ and its
left $\graa-$module \uol. A generic element $a$ of $\uol$ admits a
(fundamental) matrix representation of type
$$ a =
\left(\matrix{
0&-\beta^\diamondsuit&\beta\cr
-\beta&i k&z\cr
-\beta^\diamondsuit&-z^\diamondsuit & -i k \cr
} \right ) \qquad k \in \rr_c, z \in \graa, \beta \in \grab. $$
Also in this case, by means of the conjecture in section 3.3, we can give to
the generic $\uols$ element matrix form
$$ M_f =
\left(\matrix{
0&-\alpha^\diamondsuit&\alpha\cr
-\alpha& i \lambda&w\cr
-\alpha^\diamondsuit&-w^\diamondsuit & -i \lambda \cr} \right )
\qquad \lambda \in \rr_c, w \in \graa, \alpha \in \grab. $$
Being $\uo$ contragradient, one can obtain the orbits corresponding to coherent
states
supermanifolds from elements of the form
$$ f_0 =
\left(\matrix{
0&0&0\cr
0& i \lambda&0\cr
0&0& -i \lambda \cr} \right ) $$
stabilized by the (ordinary) Lie subgroup $U(1)$. These orbits are isomorphic
to coset space $\displaystyle{UOSP(1|2) \over U(1)}$, i.e are supermanifolds of
dimension
$(2,2)$.
We can then write the matrix for the generalized Kirillov-Kostant 2-form
\ccf, which is nondegenerate on the factor space $\alg/\algh$ (see Appendix).
Other
properties of $\ccf$ for this class of orbits are not considered here,
because irrilevant in this context.\par
\eject
\centerline{\bf 5. CONCLUSIONS}
\medskip
We have given a generalized definition of coherent states for a system
described by a hamiltonian which lives in a
dynamical superalgebra and have showed that for the SCS systems so
obtained the properties of transitivity with respect to the action of the
dynamical group, the identity resolution and overcompleteness,
as well as stability during
time evolution, still hold. For what concerns the problem of establishing
whether or not the SCS are states of minimum quantum uncertainty we
we have shown that this concept has not a well-defined, unambiguous
meaning and we can only refer to the cited literature for
some examples. \par
We have studied in detail the SCS supermanifold for three dynamical
superalgebras. In
the simplest case of $u(1|1)$ we have found an homogeneous K\"ahler
supermanifold, determined explicitly its K\"ahler structure, derived the
K\"ahler potential, and have determined the connection between the SCS
supermanifold and the coadjoint orbits. In the case of
$su(2|1)$ we have decomposed the homogeneous
supermanifold into the product of two
Einstein-K\"ahler symmetric spaces, one purely fermionic and one purely
bosonic,
for which we have determined K\"ahler structures and potentials: we expect
the product supermanifold to be K\"ahler as well.
In the case of $uosp(1|2)$ we showed that one cannot
construct a K\"ahler structure over the purely fermionic factor
supermanifold, we expect then the whole supermanifold not to be K\"ahler.
On the basis
of these results, and looking at the particular properties of the
conjugation operation that one has to introduce in the Grassmann algebra
when working with superalgebras whose IRREP can not be turned into star
representations, we conjecture that a necessary condition for our SCS
supermanifolds to be K\"ahler is that the dynamical superalgebra admits a
star representation, and that the UIR used to define the SCS system is
derived from one such IRREP. \par
In addition, we have established a connection between a certain orbit of a
generic contragradient Lie superalgebra and the coherent states
supermanifold related to non-degenerate representations of that
superalgebra. Finally, in the examples, we have found symplectic structures
over these orbits. \par
\eject
\medskip
{\bf Appendix}
\smallskip
\underbar{Inverse metric tensor for the $su(2|1)$ case.} \par
The inverse of the invariant metric tensor on $\displaystyle{SU(2|1) \over
U(1) \otimes U(1)}$ has the usual block form $g^{-1} = \pmatrix{A & B \cr C
& D \cr}$, where, in the basis $\left\{\partial_z, \partial_{z^*},
\partial_{\eta_1}, \partial_{\eta_1^*}, \partial_{\eta_2},
\partial_{\eta_2^*}\right\}$,
$$A=(1+z^*z)^2\left(1-{1 \over 4}e_1
-{1 \over 4}e_2+{1 \over 2}e_1 e_2\right)
\pmatrix{0 & 1 \cr 1 & 0 \cr},$$
$B^t = C =$
$$\left[\matrix{
-{1 \over 2}z\eta_1({\scriptstyle 1}-{1 \over 2}e_2 )
+{i \over 2}z^2\eta_2({\scriptstyle 1}-{1 \over 2}e_1 ) &
{1 \over 2}z^*\eta_1({\scriptstyle 1}-{1 \over 2}e_2 )
+{i \over 2}\eta_2({\scriptstyle 1}-{1 \over 2}e_1 ) \cr
{1 \over 2}z\eta_1^*({\scriptstyle 1}-{1 \over 2}e_2 )
-{i \over 2}\eta_2^*({\scriptstyle 1}-{1 \over 2}e_1 ) &
-{1 \over 2}z^*\eta_1^*({\scriptstyle 1}-{1 \over 2}e_2 )
-{i \over 2}{z^*}^2\eta_2^*({\scriptstyle 1}-{1 \over 2}e_1 ) \cr
{i \over 2}\eta_1({\scriptstyle 1}-{1 \over 2}e_2 )
+{1 \over 2}z\eta_2({\scriptstyle 1}-{1 \over 2}e_1 ) &
{i \over 2}{z^*}^2\eta_1({\scriptstyle 1}-{1 \over 2}e_2 )
-{1 \over 2}z^*\eta_2({\scriptstyle 1}-{1 \over 2}e_1 ) \cr
-{i \over 2}z^2\eta_1^*({\scriptstyle 1}-{1 \over 2}e_2 )
-{1 \over 2}z\eta_2^*({\scriptstyle 1}-{1 \over 2}e_1 ) &
-{i \over 2}\eta_1^*({\scriptstyle 1}-{1 \over 2}e_2 )
+{1 \over 2}z^*\eta_2^*({\scriptstyle 1}-{1 \over 2}e_1 ) \cr}\right]$$
and $D = $
$$\pmatrix{
0 & -(1 + e_2 - e_1e_2) & 0 & \eta_2^*\eta_1 \cr
1 + e_2 - e_1e_2 & 0 & -\eta_1^*\eta_2 & 0 \cr
0 & \eta_1^*\eta_2 & 0 & -(1 + e_1 - e_1e_2) \cr
-\eta_2^*\eta_1 & 0 & 1 + e_1 - e_1e_2 & 0 \cr}.
$$\par \smallskip
\par
\underbar{ $\su$'s orbit isomorphic to $ \SU/U(2) $.}
$$ \olfa = i x \pmatrix{
1-\alpha^* \alpha + \eta
&
\alpha \beta^*
&
-\alpha + \alpha \beta^* \beta
\cr
-\alpha^*\beta
&
1-\beta^*\beta+\eta
&
-\beta+\alpha^*\alpha \beta
\cr
-\alpha^*+\alpha^*\beta^*\beta
&
-\beta^*+\alpha^*\alpha \beta^*
&
2-\alpha^*\alpha-\beta^*\beta+2 \eta\cr} $$
where $\eta=\alpha^*\alpha\beta^*\beta.$\par
\smallskip
\underbar{ Generalized Kirillov-Kostant 2-form for $\su$. } \par
We list only a subset of nonzero elements, the others can be recovered by
means of superskewsymmetry
$$\eqalign{ \ccf_{1,5}=&- \ccf_{2,5} = - \ccf_{3,6} = -\alpha^*\eta_\beta \cr
\ccf_{1,7} =& \ccf_{2,7} = \ccf_{3,8} = -\beta\eta_\alpha \cr
\ccf_{2,3} =& 2 \ccf_{5,7} = -2\alpha^*\beta \cr
\ccf_{3,4} =& \beta^*\beta - \alpha^*\alpha \cr
\ccf_{4,7} =& \ccf_{1,8} = -\ccf_{2,8} = \alpha\eta_\beta \cr
\ccf_{5,8} =& 1 - \beta^*\beta\eta_\alpha \cr
\ccf_{2,4} =& 2\ccf_{6,8} = -2\alpha\beta^* \cr
\ccf_{1,6} =& \ccf_{2,6} = - \ccf_{4,5} = -\beta^*\eta_\alpha \cr
 \ccf_{6,7} =& -1 + \alpha^*\alpha\eta_\beta \cr}$$
where $\eta_\alpha = 1 - \alpha^*\alpha$, $\eta_\beta = 1 - \beta^*\beta$,
$ \ccf_{i,j}=\ccf(e_i,e_j) $ and $\{e_i, i = 1, 2, \ldots, 8\}$ is a
suitable homogeneous basis.\par
 \underbar{ Coherent states orbit for the $\uo$ case. } \par
The coordinate set of an element in generic position on an orbit
belonging to this class is
$$\eqalign{&
\{ -\xi \sin \theta \efi,-\xi \sin \theta \efis,i \xi \cos \theta \mid \cr
& \beta
\sin \theta \efi - i\beta^\diamond \cos \theta , - \beta^\diamond \sin \theta
\efis + i \beta \cos \theta \} \cr}$$
where $ \xi = 1- \beta^\diamond \beta $.\par\smallskip
\underbar{ Generalized Kirillov-Kostant 2-form for $\uo.$ } \par
We list in the following only a subset of non zero elements, the
others can be obtained by means of the skewsymmetry properties of \ccf.
$$\eqalign{
\ccf_{1,3} =& -\ccf_{5,5} = -2 \xi \sin \theta \efis \cr
\ccf_{2,3} =& \ccf_{4,4} = 2 i \xi \sin \theta  \efi \cr
\ccf_{1,5} =&  \ccf_{3,4} =
-2 \beta^\diamondsuit \sin \theta \efis + 2 i \beta \cos \theta \cr
\ccf_{2,4} =& \ccf_{3,5} =
2 \beta \sin \theta \efi - 2 i \beta^\diamondsuit \cos \theta \cr
\ccf_{4,5} =& -2i \xi \cos \theta \cr & \cr}$$
where $ \ccf_{i,j}=\ccf(e_i,e_j) $ and $\{e_i, i = 1,2, \ldots, 5\}$ is a
suitable homogeneous basis.\par
\medskip
\bf Acknowledgements \rm \par
\smallskip
We would like to thank Prof. M. Rasetti for having proposed to us this work
and for illuminating discussions. As well, we thank
R. Link, A. Montorsi and V. Penna for stimulating discussions. \par
\bigskip
\bf References \rm
\smallskip
\item{1.} R. Glauber, {\sl Phys. Rev.} {\bf 130} (1963) 2529
\item{2.} R. Glauber, {\sl Phys. Rev.} {\bf 131} (1963) 2766
\item{3.} A.O. Barut and L. Girardello, {\sl Comm. Math. Phys.} {\bf 21} (1971)
41
\item{4.} R. Delbourgo, {\sl J. Phys.} {\bf A10} (1977) 1837
\item{5.} A. M. Perelomov, {\sl Comm. Math. Phys.} {\bf 26} (1972) 222
\item{6.} M. Rasetti, G. D'Ariano and A. Montorsi, {\sl Il Nuovo Cimento}
{\bf 11D} (1989) 19
\item{7.} A.B. Balantekin, H.A. Schmitt and B.R. Barrett, {\sl J. Math.
Phys.} {\bf 29} (1988) 1634
\item{8.} A.B. Balantekin, H.A. Schmitt and P. Halse, {\sl J. Math.
Phys.} {\bf 30} (1989) 274
\item{9.} M. Chaichian, D. Ellinas and P. Presnajder, CERN preprint
CERN-TH.5925/90
\item{10.} A. Montorsi, M. Rasetti and A.I. Solomon, {\sl Int. J. Mod. Phys.}
{\bf
B3} (1989) 247
\item{11.} A. Pelizzola, {\sl Thesis} (Universit\'a di Torino, 1990,
unpublished)
\item{12.} C. Topi, {\sl Thesis} (Universit\'a di Torino, 1991, unpublished)
\item{13.} B.W. Fatyga, V.A. Kostelecky, M.M. Nieto and D.R. Truax, {\sl Phys.
Rev.} {\bf D43} (1991) 1403
\item{14.} A. Montorsi, M. Rasetti and A. I. Solomon, {\sl Phys. Rev. Lett.}
{\bf
59} (1987) 2243
\item{15.} M. Rasetti, {\sl Int. J. Theor. Phys.} {\bf 13} (1975) 425
\item{16.} L.S. Schulman, {\sl Techniques and applications of path integration}
(Wiley, New York, 1981)
\item{17.} F. A. Berezin, {\sl Introduction to superanalysis} (Reidel,
Dordrecht, 1987)
\item{18.} A. Rogers, {\sl J. Math. Phys.} {\bf 22} (1981) 939
\item{19.} F. A. Berezin, {\sl The method of second quantization} (Academic
Press,
New York, 1966)
\item{20.} V. G. Kac, {\sl Comm. Math. Phys.} {\bf 53} (1977) 31
\item{21.} R. Delbourgo and J.R. Fox, {\sl J. Phys.} {\bf A10} (1977) L233
\item{22.} A. M. Perelomov, {\sl Generalized coherent states and their
applications}
\item{} (Springer-Verlag, Berlin, 1986)
\item{23.} W. Magnus, {\sl Comm. Pure Appl. Math.} {\bf VII} (1954) 649
\item{24.} G. D'Ariano, M. Rasetti and M. Vadacchino, {\sl J. Phys.} {\bf A18}
(1985) 1295
\item{25.} M. Bordemann, M. Forger and H. R\"omer, {\sl Comm. Math. Phys.} {\bf
102} (1986) 605
\item{26.} A. Cavalli, G. D'Ariano and L. Michel, {\sl Ann. Inst. H.
Poincar\'e}
{\bf 44} (1986) 173
\item{27.} H. Kuratsuji and T. Suzuki, {\sl Progr. Theor. Phys. Suppl.} {\bf
74-75} (1983) 209
\item{28.} G. Giavarini and E. Onofri, {\sl J. Math. Phys.} {\bf 30} (1989)
659
\item{29.} M.V. Berry, {\sl Proc. Roy. Soc. London} {\bf A392} (1984) 45
\item{30.} A. Montorsi and M. Rasetti, private communication (see also Ref.
33)
\item{31.} J. Hubbard, {\sl Proc. Roy. Soc. London} {\bf A276} (1963) 238
\item{32.} G.V. Dunne, R. Jackiw and C.A. Trugenberger, {\sl Ann. Phys.}
{\bf 194} (1989) 197
\item{33.} A. Montorsi, M. Rasetti, {\sl Mod. Phys. Lett.} {\bf B4} (1990) 613
\item{34.} M. Scheunert, W. Nahm and V.Rittenberg, {\sl J. Math. Phys.} {\bf
18} (1977)
146
\item{35.} B. De Witt, {\sl Supermanifolds} (Cambridge University Press,
Cambridge, 1984)
\item{36.} A. Kirillov, \sl Russian Math. Surveys \bf 17 \rm (1962) 53
\item{37.} A. Kirillov, \sl Elements of the theory of representations; \it
 Grundlehren der mathematischen Wissenschaften n. \bf
 220 \rm (Springer-Verlag, Berlin, 1976)
\item{38.} H. Moscovici, \sl Comm. Math. Phys. \bf 54 \rm (1977) 63
\item{39.} A. Coleman, \sl Induced and subdued representations \rm in
\it Group theory and its applications \rm Ed. E.M. Loebl (Academic Press,
New York, 1968)
\item{40.} G.W. Mackey, {\sl Ann. of Math.} {\bf 55} (1952) 101
\item{41.} G.W. Mackey, {\sl Ann. of Math.} {\bf 58} (1953) 193
\item{42.} B. Kostant, \sl Quantization and unitary representations. Part I.
Prequantization. \it Lectures in Modern Analysis and Applications, III. \rm
 Lecture Notes in Mathematics Vol \bf 170, \rm  pag 87-208
(Springer-Verlag, Berlin, 1970)
\item{43.} L. Auslander and B. Kostant, \sl Bull. Am. Math. Soc. \bf 73 \rm
(1967) 692
\item{44.} M. Scheunert, W. Nahm and V. Rittenberg, {\sl J. Math. Phys.} {\bf
18} (1977)
155
\item{45.} V. Rittenberg, {\sl A guide to Lie superalgebras}, {\rm in} {\it
Proc. of the VI International Colloquium on Group Theoretical
Methods in Physics} (T\"ubingen, 1977)
\item{46.} F.A. Berezin and V. Tolstoy, \sl Comm. Math. Phys. \bf 78 \rm
(1981) 409
\end